\documentclass[preprint2]{aastex}

\newcommand{\Hb} {H$\beta$}
\newcommand{\Ha} {H$\alpha$}
\newcommand{\Hg} {H$\gamma$}
\newcommand{\Hd} {H$\delta$}
\newcommand{\He} {H$\epsilon$}
\begin{document}

\title{The localized chemical pollution in NGC~5253 revisited: \\ Results from deep echelle spectrophotometry\footnotemark{}}

\author{\'Angel R. L\'opez-S\'anchez}
\affil{Instituto de Astrof{\'\i}sica de Canarias, E-38200, La Laguna, Tenerife, Spain, angelrls@iac.es}
\author{C\'esar Esteban}
\affil{Instituto de Astrof{\'\i}sica de Canarias, E-38200, La Laguna, Tenerife, Spain, cel@iac.es}
\author{Jorge Garc\'{\i}a-Rojas}
\affil{Instituto de Astrof{\'\i}sica de Canarias, E-38200, La Laguna, Tenerife, Spain, jogarcia@iac.es}
\author{Manuel Peimbert}
\affil{Instituto de Astronom\'\i a, UNAM, 
Apdo. Postal 70-264, M\'exico 04510 D.F., Mexico}
\email{peimbert@astroscu.unam.mx}
\and
\author{M\'onica Rodr\'{\i}guez}
\affil{Instituto Nacional de Astrof\'{\i}sica, \'Optica y Electr\'onica, Apdo. Postal 51 y 216, 72000 Puebla, Mexico}
\email{mrodri@inaoep.mx} 

\begin{abstract}
We present echelle spectrophotometry of the blue compact dwarf galaxy (BCDG) NGC~5253. The data have been taken with the Very Large Telescope UVES 
echelle spectrograph in the 3100 to 10400 \AA\ range. We have measured the intensities of a large number of permitted and forbidden emission lines in 
four zones of the central part of the galaxy. In particular, we detect faint \ion{C}{2} and \ion{O}{2} recombination lines. This is the first time 
that these lines are unambiguously detected in a dwarf starburst galaxy.  The physical conditions of the ionized gas have been derived using a large 
number of different line intensity ratios. Chemical abundances of He, N, O, Ne, S, Cl, Ar, and Fe have been determined following the standard 
methods. In addition, C$^{++}$ and O$^{++}$ abundances have been derived from pure recombination lines. These abundances are larger than those 
obtained from collisionally excited lines (from 0.30 to 0.40 dex for C$^{++}$ and from 0.19 to 0.28 dex for O$^{++}$). This result is consistent with 
a temperature fluctuations parameter ($t^2$) between 0.050 and 0.072. We confirm previous results that indicate the presence of a localized N 
enrichment in certain zones of the center of the galaxy. Moreover, our results also indicate a possible slight He overabundance in the same zones. 
The enrichment pattern agrees with that expected for the pollution by the ejecta of massive stars in the Wolf-Rayet (WR) phase. The amount of 
enriched material needed to produce the observed overabundance is consistent with the mass lost by the number of WR stars estimated in the 
starbursts. Finally, we discuss the possible origin of the difference between abundances derived from recombination and collisionally excited lines 
(the so-called {\it abundance discrepancy} problem) in \ion{H}{2} regions, finding that a recent hypothesis based on the delayed enrichment by SNe 
ejecta inclusions seems not to explain the observed features.  
\end{abstract}

\keywords{galaxies: starburst --- galaxies: abundances --- galaxies:  kinematics and dynamics --- 
galaxies: clusters: individual: NGC~5253}

\section{Introduction}

\footnotetext{Based on observations collected at the European Southern Observatory, Chile, proposal number ESO 
70.C-0008(A)}

Deep cross-dispersed echelle spectrophotometry with large aperture telescopes is a powerful technique for refining 
our knowledge about the che\-mical composition of \ion{H}{2} regions. It permits to observe the whole optical spectral 
range, to measure crucial faint auroral and recombination lines, to deblend nebular lines, and to decontaminate them 
for nearby sky spectral features. 

Our group has obtained deep high resolution spectra of most of the brightest Galactic \ion{H}{2} regions 
\citep{E98,E99a,E99b,E04,GRE04,GRE05,GRE06} that have led to the precise determination of the abundance of heavy 
element ions from recombination lines and even to the determination of the carbon radial abundance gradient in the 
Galactic disk \citep{E05}. There is still a limited number of similar stu\-dies devoted to giant extragalactic 
\ion{H}{2} regions (hereafter GEHRs), focusing the efforts on the Magellanic Clouds \citep{P03,Tsamis03} and in a few 
other nearby irregular and spiral galaxies \citep{EP02,PPR05}. A common result of both the Galactic and extragalactic 
works, is the finding of a systematic difference between the abundances determined from recombination lines 
(hereafter RLs) and collisionally excited lines (hereafter CELs) of the same ions, in the sense that abundances 
determined from RLs are {\it always} higher than those determined from CELs. This problem, that has been also 
reported to occur in some planetary ne\-bulae (PNe), is known as the {\it abundance discre\-pancy}. The origin of this 
problem is still unknown. \citet{Pequignot02} and \citet{Tsamis04} have constructed models with inhomogeneous 
chemical composition and physical conditions to explain that discre\-pancy in PNe. A similar idea has been very recently proposed by 
\citet{Tsamis05} for explaining the abundance discrepancy in the case of 30 Dor. These authors propose the presence of
metal enhanced inclusions that would be responsible for most of the emission in RLs of heavy element ions. 
However, the abundance discrepancy could be related to another secular problem in nebular 
astrophysics: the existence of temperature fluctuations in the ionized gas volume (Peimbert 1967; Torres-Peimbert, Peimber \& Daltabuit 1980),
whose existence 
is still controversial although some mechanisms have been proposed (see Esteban 2002 for a review). 

\begin{figure*}[t!]
\includegraphics[width=1\linewidth]{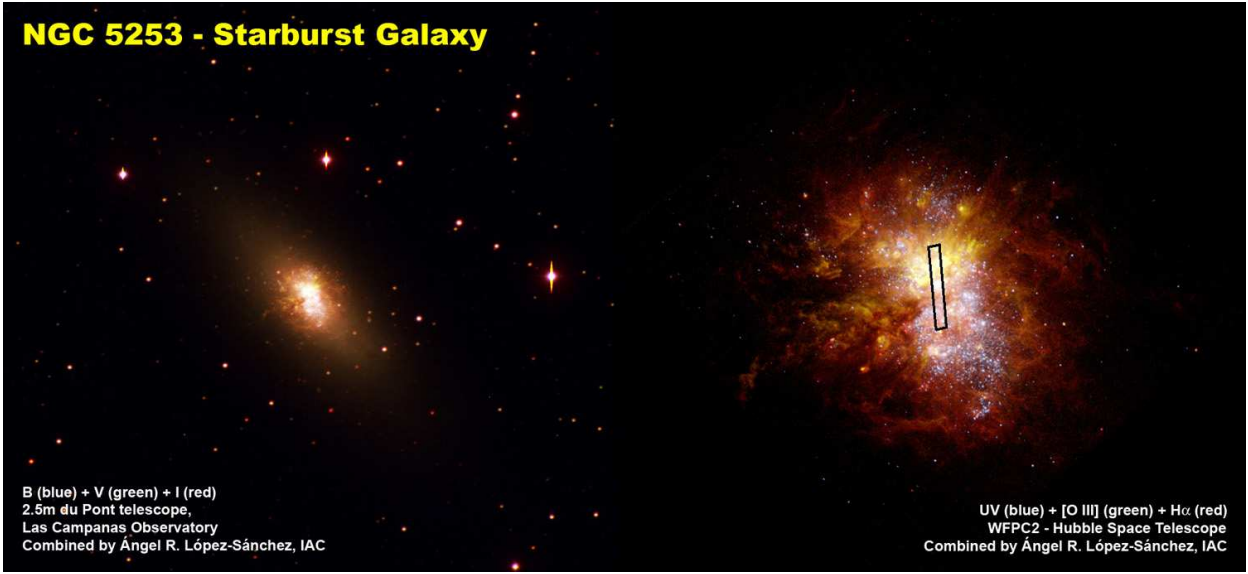}
\caption{\small{(\emph{Left}) False color image in filters $B$ (blue), $V$ (green) and $I$ (red) of the starburst galaxy NGC 
5253 observed by the 2.5m du Pont telescope, Las Campanas Observatory. (\emph{Right}) False color image obtained by  
combining $HST$ \Ha\ (red), [O III] (green) and UV (blue) images. The position of our $VLT$ slit is indicated in it. North is up and east at left. 
The $HST$ image has been included at the center of the 2.5m du Pont telescope image for comparison.}}
\label{nombres}
\end{figure*}

One step ahead in our investigation is to obtain abundance determinations from RLs in a sample of bright GEHRs in 
starburst galaxies and even \ion{H}{2} galaxies of chemical compositions other than solar. The dwarf irregular galaxy 
NGC 5253 is one of the closest starbursts and a very appro\-priate target for our purposes (Figure~\ref{nombres}). It lies at a heliocentric 
distance of 3.3 Mpc \citep{Gibson00} and belongs to the Centaurus Group. Consequently, NGC 5253 has been one of the 
most studied starbursts, observed at basically all wavelengths from radio to X-rays. 

NGC 5253 is also especially interesting because it is the best candidate for localized chemical enrichment in a GEHR. 
\citet{Welch70}, Walsh \& Roy (1987, 1989), and \citet{Kobulnicky97} reported the presence of a strong nitrogen overabundance in a 
particular zone of the starbursting nucleus of the galaxy. \citet{CTM86} also found a helium enhancement in that 
zone, but this has not been confirmed in later works. Another peculiarity of the galaxy is that the
majority of the gas seems to rotate about the optical major axis of the galaxy \citep{KS95} but this behaviour is not
clear and could be combined with some kind of outflow 
(Koribalski 2006, priv. communication). However, CO observations by \citet*{TBH97} and \citet*{MTB02} 
suggest that mole\-cular clouds are infalling into NGC 5253. These two observational facts indicate that the dynamical 
situation of this galaxy is far from clear. Other authors as \citet{vdBergh80} and 
\citet{CP89} have suggested that the nuclear starburst was triggered as a result of a past interaction (around 1 Gyr 
ago) with the neighboring galaxy M83. Both galaxies lie at a radial distance of only 500 kpc and the projected 
separation is 130 kpc \citep{Thim03}.

Concerning the kinematics of the ionized gas of NGC 5253, it has to be highlighted the presence of two large 
superbubbles related to the central region of the galaxy, with diameters of the order of 1 arcmin ($\sim$1 kpc), and expansion 
velocities of 35 km s$^{-1}$ \citep{Marlowe95}. These superbubbles coincide with extended X-ray emission as 
demonstrated by {\it Chandra} and {\it XMM-Newton} data analyzed by \citet{Summers04}. 
The structure and morphology of the ionized gas in NGC 5253 were deeply analyzed by \citet{Calzetti04} \mbox{using} imagery from {\it Hubble Space 
Telescope} (HST), detecting faint arches and filaments in both \Ha\ and [\ion{S}{2}] at $\geq$1 kpc from the main ionizing cluster (see 
Figure~\ref{nombres}, right
panel) that are partially excited by shocks. 

NGC 5253 has been considered as one of the youngest starbursts in the local universe \citep{vdBergh80,MG82,Rieke88}.
\citet*{CTM86} and \citet{WR87} 
reported a broad emission feature, indicating the presence of Wolf-Rayet (WR) stars in this galaxy. 
Seve\-ral authors \citep{Schaerer97,Kobulnicky97} have confirmed the presence of late-type WN and early-type WC stars. 
This fact, and the almost entirely thermal radio spectrum with very little synchrotron emission from supernova 
remants \citep{Beck96}, imply the extreme youth of the starburst. Indeed, there is evidence in the literature that multiple starbursts 
in succession have occurred in the center of this galaxy, leaving behind an evolved, but still young-ish stellar population \citep{SS00,Tremonti01}.

\section{Observations and data reduction}

The observations were made on 2003 March 30 with the Ultraviolet Visual Echelle Spectrograph, UVES \citep{dodo00}, at 
the {\it Very Large Telescope}, VLT, Kueyen unit in Cerro Paranal Observatory (Chile). The standard settings in both 
the red and blue arms of the spectrograph, covering the region from 3100 to 10400 \AA, were used. The wavelength 
intervals 5783--5830 \AA\ and 8540--8650 \AA\ were not observed due to a gap between the two CCDs used in the red arm. 
There are also five small intervals that were not observed, 9608--9612 \AA, 9761--9767 \AA, 9918--9927 \AA, 
10080--10093 \AA\ and 10249--10264 \AA, because the five redmost orders did not fit completely within the CCD. The 
whole spectrum was taken in two different blocks of observations. The first block covers 3800-5000 \AA\ with the blue 
arm and 6700--10400 \AA\ with the red arm and was taken in three consecutive 1000s exposures. The second block, 
covering 3100--3900 \AA\ and 4750--6800 \AA, was taken in two consecutive 360s exposures. None of the emission lines 
was saturated in the exposures. The journal of the observations can be found in Table~\ref{observa}.

\begin{table}[t!]\centering
  \caption{Journal of observations.}
  \smallskip
  \label{observa}  
  \small
  \begin{tabular}{ccccc}
    \hline\hline
	\noalign{\smallskip}
    $\Delta\lambda$  & Exp. time  & Spectral res.$\rm^a$   & Spatial res. \\ 
        (\AA)  & (s)  & (\AA\ pix$^{-1}$) &  ($\arcsec$ pix$^{-1}$) \\
	\hline
	\noalign{\smallskip}
	3100--3900 &  2 x 360  & 0.019  & 0.25     \\
	3800--5000 &  3 x 1000 & 0.022  & 0.25    \\
    4750--6800 &  2 x 360  & 0.024  & 0.18    \\
   6700--10400 &  3 x 1000 & 0.033  & 0.17  \\
    \noalign{\smallskip}
	\hline\hline
  \end{tabular}
   \begin{flushleft}
  $\rm^a$ At the center of each wavelength range.
  \end{flushleft}
\end{table}

The atmospheric dispersion corrector (ADC) was used to keep the same observed region within the slit regardless of 
the air mass value. This is especially important for this work because we extracted, analyzed and compared different 
small areas along the slit. The slit width was set to 1.5$\arcsec$ and the slit length was set to 10$\arcsec$ in the 
blue arm and to 12$\arcsec$ in the red arm.  The slit width was chosen to maximize the S/N ratio of the
emission lines, to separate most of the relevant faint lines, and to obtain a good spectral resolution for ana\-lyzing 
the velocity structure of the ionized gas. The effective resolution at a given wavelength is approximately $\Delta 
\lambda \sim \lambda / 17600$. The seeing was excellent during the observations, $\sim$0.5 $\arcsec$.

\begin{figure}[t!]
\includegraphics[width=1\linewidth]{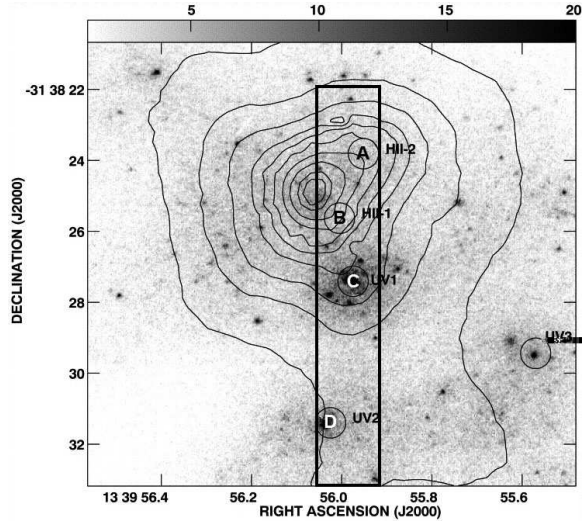}
\caption{\small{Our VLT slit position over Figure 1 of \citet{Kobulnicky97}, that shows the \Ha\ contours over the 
2200 \AA\ continuum image (gray scale). The four analyzed knots inside NGC 5253 are indicated and labeled.}}
\label{kobul}
\end{figure}

In Figure~\ref{kobul} we show our slit position over the Figure 1 of \citet{Kobulnicky97}. The slit position was 
located along the north-south direction  (P.A. = 0${^\circ}$), which was chosen in order to observe the most 
interesting regions inside the starburst. These zones were previously analyzed by \citet{WR89} and 
\citet{Kobulnicky97}. From north to south, these regions were designated HII-2, HII-1, UV1, and UV2 (A, B, C, and D, respectively), 
as it is shown in Figure~\ref{kobul}. We have extracted 1-D spectra of the regions, with a size of 1.5$\arcsec \times 
1.5\arcsec$. 

\begin{figure*}[t]
\includegraphics[angle=270,width=0.92\linewidth]{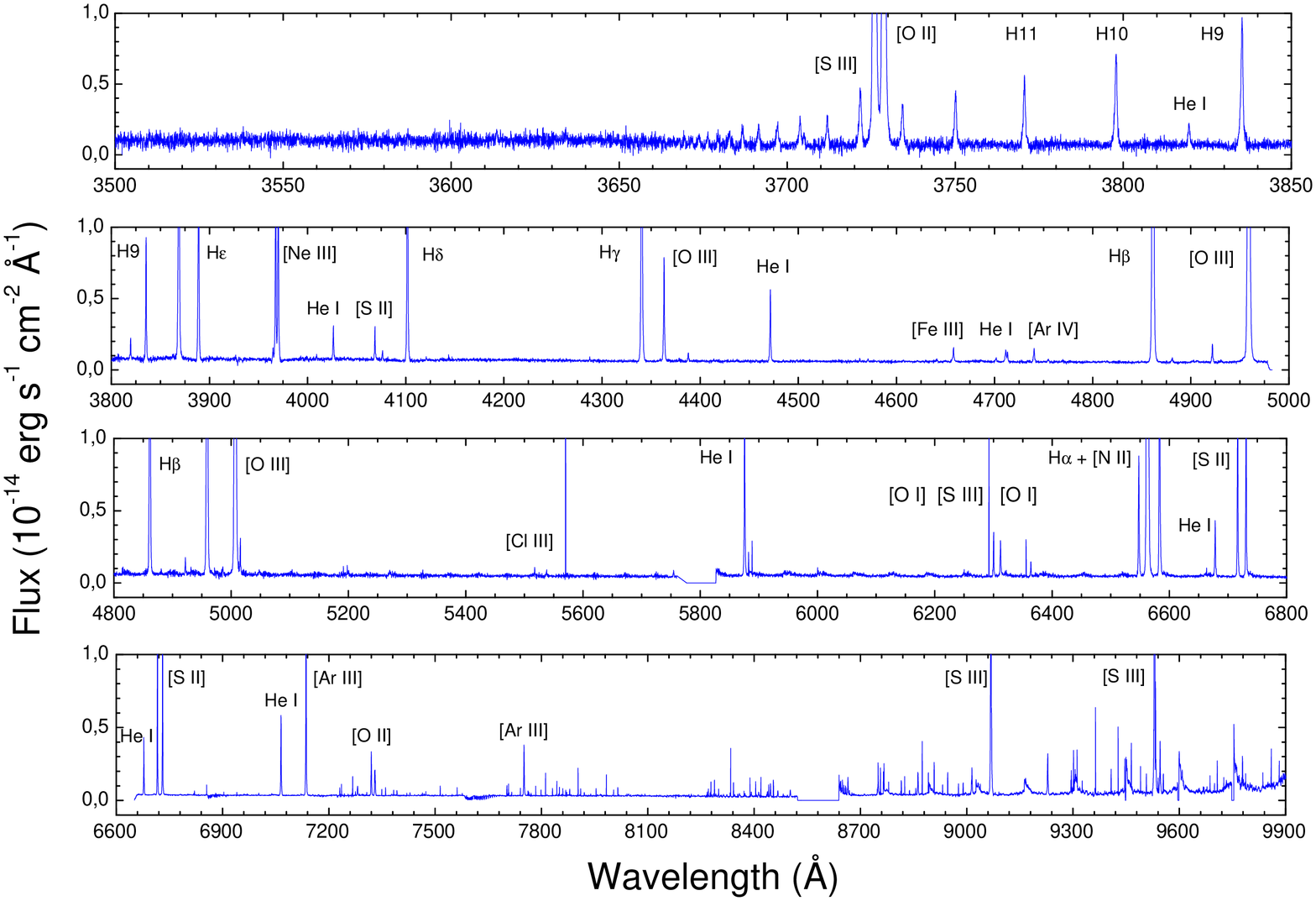}
\caption{\small{VLT UVES spectra of region B (HII-1). Some of the brightest emission lines are labeled. Telluric lines have not been removed and are 
specially evident in the 6600--9000 \AA\ range. Note the gaps in the intervals 5783--5830 \AA\ and 8540--8650 \AA.}}
\label{spectra}
\end{figure*}

The spectra were reduced using the IRAF\footnote{IRAF is distributed by NOAO which is operated by AURA Inc., under 
cooperative agreement with NSF} echelle reduction package, following the standard procedure of bias subtraction, 
flatfielding, wavelength calibration, aperture extraction and flux calibration. 
The standard stars EG 247, HD 49798, and C-32d9927 were observed for flux calibration. As an example, we show the wavelength- and 
flux-calibrated spectra of region B (HII-1) in Figure~\ref{spectra}.

\section{Line intensities \\ and reddening correction}

Line intensities and equivalent widths were measured integrating all the flux in the line between two given limits and
over a local continuum estimated by eye. In the cases of line blending, a multiple Gaussian profile fit procedure was 
applied to obtain the line flux of each individual line. The measurements were performed with the SPLOT routine of 
the IRAF package.

In Table~\ref{lineid} we have compiled all the emission lines detected in the four analyzed regions. Knot B (HII-1) is the 
zone where more lines have been identified: 169. Therefore, NGC 5253 is nowadays the dwarf starburst galaxy where more 
optical emission lines have been reported. We identify 156 lines in A (HII-2), 161 in C 
(UV1), and 86 in D (UV2). Besides the \ion{H}{1} Paschen lines included in Table~\ref{lineid}, some other lines were 
also detected, but they are blended with telluric spectral features, making it impossible to obtain a good determi\-nation of their 
intensity. The identification and adopted laboratory wavelength of the lines, as well as their errors, were obtained following \citet{GRE04,E04}. 
Colons indicate errors of the order or greater than 40\%.

\begin{deluxetable}{c@{\hspace{5pt}}c@{\hspace{5pt}}cccccc} 
\tabletypesize{\scriptsize}
\tablecaption{Dereddened line intensity ratios with respect to I(H$\beta$)=100.
\label{lineid}}
\tablewidth{0pt}
\tablehead{
\colhead{$\lambda_0$ (\AA)} &
\colhead{Ion} &
\colhead{Mult.} & 
\colhead{$f(\lambda)$}  &
\colhead{(A) HII-2} &
\colhead{(B) HII-1} &
\colhead{(C) UV-1} &
\colhead{(D) UV-2}} 
\startdata
 3187.84 &  He I &    3F &  0.308 &  4.01 $\pm$ 0.33   &  2.89 $\pm$ 0.28   &  3.82 $\pm$ 0.26   &   \nodata\\
 3447.49 &  He I &     7 &  0.295 &        \nodata   &      0.309:   &        \nodata   &        \nodata\\
 3450.39 & [Fe II] &   27F &  0.295 &        \nodata   &      0.063:   &        \nodata   &        \nodata\\

 3530.50 &  He I &    36 &  0.286 &        \nodata   &      0.152:   &        \nodata   &        \nodata\\
 3554.42 &  He I &    34 &  0.283 &      0.249:   &      0.158:   &  0.44 $\pm$ 0.11   &        \nodata\\
 3587.28 &  He I &    32 &  0.278 &      0.322:   &      0.191:   &        \nodata   &        \nodata\\
 3613.64 &  He I &     6 &  0.275 &      0.335:   &      0.356:   &  0.34 $\pm$ 0.10   &        \nodata\\
 3634.25 &  He I &    28 &  0.272 &      0.313:   &      0.411:   &  0.28 $\pm$ 0.10   &        \nodata\\
 3651.97 &  He I &    27 &  0.269 &        \nodata   &      0.093:   &        \nodata   &        \nodata\\
 3664.68 &   H I &   H28 &  0.267 &        \nodata   &        \nodata   & 0.24 $\pm$ 0.09   &        \nodata\\
 3666.10 &   H I &   H27 &  0.267 &      0.159:   &      0.124:   &  0.35 $\pm$ 0.10   &        \nodata\\
 3667.28 &   H I &   H26 &  0.266 &      0.130:   &      0.144:   & 0.276 $\pm$ 0.097   &        \nodata\\
 3669.47 &   H I &   H25 &  0.266 &      0.172:   &      0.183:   & 0.271 $\pm$ 0.097   &        \nodata\\
 3671.48 &   H I &   H24 &  0.266 &      0.166:   &      0.205:   & 0.256 $\pm$ 0.095   &        \nodata\\
 3673.76 &   H I &   H23 &  0.265 &      0.187:   &      0.255:   &  0.51 $\pm$ 0.12   &        \nodata\\
 3676.37 &   H I &   H22 &  0.265 &      0.284:   &      0.316:   &  0.51 $\pm$ 0.12   &        \nodata\\
 3679.36 &   H I &   H21 &  0.265 &      0.374:   &      0.229:   &  0.57 $\pm$ 0.12   &        \nodata\\
 3682.81 &   H I &   H20 &  0.264 &      0.365:   &      0.389:   &  0.68 $\pm$ 0.13   &        \nodata\\
 3686.83 &   H I &   H19 &  0.263 &  0.68 $\pm$ 0.20   &  0.56 $\pm$ 0.18   &  0.71 $\pm$ 0.13   &    \nodata\\
 3691.56 &   H I &   H18 &  0.263 &  0.91 $\pm$ 0.22   &  0.76 $\pm$ 0.19   &  0.97 $\pm$ 0.15   &      0.570:\\
 3697.15 &   H I &   H17 &  0.262 &  1.18 $\pm$ 0.23   &  1.01 $\pm$ 0.21   &  1.21 $\pm$ 0.16   &      1.021:\\
 3703.86 &   H I &   H16 &  0.260 &  1.30 $\pm$ 0.23   &  1.26 $\pm$ 0.22   &  1.24 $\pm$ 0.16   &      0.736:\\
 3705.04 &  He I &    25 &  0.260 &  0.58 $\pm$ 0.19   &      0.414:      &  0.49 $\pm$ 0.12   &        \nodata\\
 3711.97 &   H I &   H15 &  0.259 &  1.40 $\pm$ 0.24   &  1.27 $\pm $0.22   &  1.41 $\pm$ 0.17   &      1.324:\\
 3721.83 & [S III] &  2F &  0.257 &  2.95 $\pm$ 0.29   &  3.12 $\pm$ 0.28   &  3.24 $\pm$ 0.24   &  2.96 $\pm$ 0.69\\
 3726.03 & [O II] &    1F &  0.257 &  65.3 $\pm$ 2.3   &  62.9 $\pm$ 2.2   &  80.3 $\pm$ 2.9   & 145.4 $\pm$ 5.3\\
 3728.82 & [O II] &    1F &  0.256 &  71.9 $\pm$ 2.6   &  71.9 $\pm$ 2.6   & 102.2 $\pm$ 3.6   & 208.0 $\pm$ 7.2\\
 3734.17 &  H I &   H13 &  0.255 &  2.62 $\pm$ 0.28   &  2.16 $\pm$ 0.25   &  2.31 $\pm$ 0.20   &  2.60 $\pm$ 0.67\\
 3750.15 &  H I &   H12 &  0.253 &  3.14 $\pm$ 0.30   &  2.89 $\pm$ 0.27   &  2.80 $\pm$ 0.22   &  2.78 $\pm$ 0.68\\
 3770.63 &  H I &   H11 &  0.249 &  3.96 $\pm$ 0.32   &  3.62 $\pm$ 0.29   &  3.90 $\pm$ 0.26   &  4.59 $\pm$ 0.80\\
 3797.90 &  H I &   H10 &  0.244 &  5.32 $\pm$ 0.36   &  5.08 $\pm$ 0.33   &  5.13 $\pm$ 0.30   &  5.33 $\pm$ 0.84\\
 3819.61 &  He I &    22 &  0.240 &  1.10 $\pm$ 0.22   &  0.93 $\pm$ 0.20   &  0.94 $\pm$ 0.14   &        \nodata\\
 3835.39 &  H I &    H9 &  0.237 &  7.39 $\pm$ 0.28   &  7.37 $\pm$ 0.27   &  7.44 $\pm$ 0.28   &  7.27 $\pm$ 0.48\\
 3856.02 &  Si II &    1 &  0.233 & 0.079 $\pm$ 0.029  & 0.204 $\pm$ 0.031  & 0.086 $\pm$ 0.031  &        \nodata\\
 3862.59 &  Si II &    1 &  0.232 & 0.073 $\pm$ 0.029  & 0.084 $\pm$ 0.022  & 0.109 $\pm$ 0.034  &  0.52 $\pm$ 0.17\\
 3868.75 & [Ne III] & 1F &  0.230 &  48.8 $\pm$ 1.7    &  47.9 $\pm$ 1.6    &  34.1 $\pm$ 1.2    &  24.5 $\pm$ 1.0\\
 3871.82 &  He I &    60 &  0.230 &        \nodata     & 0.092 $\pm$ 0.023  &      0.038:        &        \nodata\\
 3889.05 &   H I &    H8 &  0.226 & 19.19 $\pm$ 0.67   & 17.91 $\pm$ 0.63   & 18.15 $\pm$ 0.64   & 18.67 $\pm$ 0.86\\
 3926.53 &  He I &    58 &  0.219 &        \nodata     & 0.070 $\pm$ 0.021  & 0.135 $\pm$ 0.036  &        \nodata\\
 3964.73 &  He I &     5 &  0.211 & 0.714 $\pm$ 0.063  & 0.680 $\pm$ 0.052  & 0.637 $\pm$0.063   &      0.336:\\
 3967.46 & [Ne III] & 1F &  0.210 & 13.81 $\pm$ 0.48   & 14.40 $\pm$ 0.50   &  9.51 $\pm$ 0.35   &  6.24 $\pm$ 0.46\\
 3970.07 &   H I &   H7 &  0.210 & 15.98 $\pm$ 0.55   & 16.03 $\pm$ 0.55   & 16.10 $\pm$ 0.56   & 15.83 $\pm$ 0.75\\
 4009.22 &  He I &    55 &  0.202 & 0.134 $\pm$ 0.034  & 0.232 $\pm$ 0.032  & 0.130 $\pm$ 0.035  &        \nodata\\
 4026.21 &  He I &    18 &  0.198 &  1.93 $\pm$ 0.10   & 1.815 $\pm$ 0.090  &  1.84 $\pm$ 0.10   &  1.51 $\pm$ 0.25\\
 4068.60 & [S II] &  1F &  0.189 &  2.37 $\pm$ 0.12   & 1.602 $\pm$ 0.082  & 1.484 $\pm$ 0.091  &  3.46 $\pm$ 0.35\\
 4069.62 &  O II &    10 &  0.189 &        \nodata   &        \nodata   &      0.054:   &        \nodata\\
 4072.15 &  O II &    10 &  0.188 &        \nodata   &        \nodata   &      0.020:   &        \nodata\\
 4076.35 & [S II] &  1F &  0.187 & 0.757 $\pm$ 0.064   & 0.490 $\pm$ 0.044   & 0.471 $\pm$ 0.055 & 1.17 $\pm$ 0.23\\
 4097.26 &   O II & 20-48 &  0.183 &      0.031:   &        \nodata   &        \nodata   &        \nodata\\
 4101.74 &   H I &    H6 &  0.182 & 26.00 $\pm$ 0.86   & 26.20 $\pm$ 0.86   & 26.05 $\pm$ 0.86   &  25.9 $\pm$ 1.0\\
 4120.82 &  He I &  16 &  0.177 & 0.152 $\pm$ 0.036   & 0.188 $\pm$ 0.030   & 0.094 $\pm$ 0.031   &        \nodata\\
 4143.76 &  He I &  53 &  0.172 & 0.212 $\pm$ 0.040   & 0.209 $\pm$ 0.031   & 0.199 $\pm$ 0.040   &        \nodata\\
 4168.97 &  He I &    52 &  0.167 &        \nodata   &      0.034:   &      0.039:   &        \nodata\\
 4243.97 & [Fe II] &   21F &  0.149 & 0.068 $\pm$ 0.027   & 0.069 $\pm$ 0.020   &      0.037:   &      0.124:\\
 4267.15 &   C II &     6 &  0.144 &  0.063 $\pm$ 0.026   & 0.074:   & 0.062 $\pm$ 0.026   &      \nodata\\
 4276.83 & [Fe II] &  21F &  0.142 &      0.023:   &      0.029:   &      0.019:   &        \nodata\\
 4287.40 & [Fe II] & 7F &  0.139 & 0.143 $\pm$ 0.034   & 0.108 $\pm$ 0.023   & 0.184 $\pm$ 0.039  &  0.40 $\pm$0.15\\
 4303.61 &   O II &    66 &  0.135 &        \nodata   &        \nodata   &      0.025:   &        \nodata\\
 4340.47 &   H I &  H$\gamma$ &  0.127 &  47.0 $\pm$ 1.5   &  46.8 $\pm$ 1.5   &  46.9 $\pm$ 1.5  &  46.9 $\pm$ 1.7\\
 4359.34 & [Fe II] &  7F &  0.122 &      0.067:   & 0.103 $\pm$ 0.023   & 0.099 $\pm$ 0.031   &      0.279:\\
 4363.21 & [O III] &  2F &  0.121 &  6.46 $\pm$ 0.23   &  6.70 $\pm$ 0.23   &  3.95 $\pm$ 0.16   &  2.61 $\pm$ 0.31\\
 4368.25 &   O I &     5 &  0.120 &      0.019:   & 0.068 $\pm$ 0.019   &      0.037:   &        \nodata\\
 4387.93 &  He I &    51 &  0.115 & 0.419 $\pm$ 0.049   & 0.474 $\pm$ 0.042   & 0.403 $\pm$ 0.050   &      0.322:\\
 4413.78 & [Fe II] & 7F &  0.109 &        \nodata   & 0.095 $\pm$ 0.022   & 0.073 $\pm$ 0.028   &        \nodata\\
 4416.27 & [Fe II] &    6F &  0.109 &      0.052:   & 0.045 $\pm$ 0.017   &      0.043:   &        \nodata\\
 4437.55 &  He I &    50 &  0.104 &      0.052:   & 0.044 $\pm$ 0.016   &      0.046:   &        \nodata\\
 4452.11 & [Fe II] &    7F &  0.100 &      0.049:   &  4.20 $\pm$ 0.16   &      0.032:   &        \nodata\\
 4471.48 &  He I &  14 &  0.095 &  4.11 $\pm$ 0.16   &  4.09 $\pm$ 0.15   &  3.79 $\pm$ 0.15   &  3.36 $\pm$ 0.34\\
 4562.60 &  Mg I] &  1 &  0.073 & 0.152 $\pm$ 0.034   & 0.108$\pm$ 0.022   & 0.129 $\pm$ 0.033   &  0.49 $\pm$ 0.16\\
 4571.20 &  Mg I] &  1 &  0.071 & 0.139 $\pm$ 0.033   & 0.102$\pm$ 0.022   & 0.115 $\pm$ 0.032   &      0.290:\\
 4638.86 &  O II &   1 &  0.055 &      0.040:   &        \nodata   &        0.049:   &        \nodata\\
 4641.81 &  O II &   1 &  0.054 &      0.066:   &        \nodata   &        0.060 $\pm$ 0.019   &        \nodata\\
 4649.13 &  O II &   1 &  0.052 & 0.063 $\pm$ 0.020   &  0.063 $\pm$ 0.021   & 0.062 $\pm$ 0.019 &    \nodata\\
 4650.84 &  O II &   1 &  0.052 & 0.034:	   &  0.036:		   & 0.042:	 &        \nodata\\
 4658.10 & [Fe III]&  3F &  0.050 & 1.185 $\pm$ 0.074   & 1.134 $\pm$ 0.062  & 0.815$\pm$ 0.065 &  2.11 $\pm$ 0.28\\
 4661.63 &   O II &     1 &  0.049 &    0.027:   &    0.026:   &      0.061: &        \nodata\\
 4701.53 & [Fe III] &  3F &  0.039 & 0.291 $\pm$ 0.042   & 0.339 $\pm$ 0.035   & 0.195 $\pm$ 0.038   &      0.340:\\
 4711.37 & [Ar IV] &  1F &  0.037 & 0.681 $\pm$ 0.058   & 0.890 $\pm$ 0.054   & 0.170 $\pm$ 0.036   &    \nodata\\
 4713.14 &  He I &  12 &  0.037 & 0.578 $\pm$ 0.054   & 0.587 $\pm$ 0.044   & 0.396 $\pm$ 0.049   &        \nodata\\
 4733.93 & [Fe III] & 3F &  0.031 & 0.090 $\pm$ 0.029   & 0.093 $\pm$ 0.021   &      0.048:   &        \nodata\\
 4740.16 & [Ar IV] & 1F &  0.030 & 0.723 $\pm$ 0.059   & 0.918 $\pm$ 0.055   & 0.135 $\pm$ 0.033   &     \nodata\\
 4754.83 & [Fe III] & 3F &  0.026 & 0.176 $\pm$ 0.035   & 0.157 $\pm$ 0.025   & 0.102 $\pm$ 0.030   &   0.240:\\
 4769.60 & [Fe III] &    3F &  0.023 & 0.131 $\pm$ 0.032   & 0.072 $\pm$ 0.019   &      0.058:   &    \nodata\\
 4814.55 & [Fe  II] &   20F &  0.012 &      0.055:         & 0.044 $\pm$ 0.016   &      0.039:   &    \nodata\\
 4861.33 &  H I &  \Hb &  0.000 & 100.0 $\pm$ 3.0     & 100.0 $\pm$ 3.0     & 100.0 $\pm$ 3.0     & 100.0 $\pm$ 3.2\\
 4881.00 & [Fe III] &  2F & -0.005 & 0.345 $\pm$ 0.044   & 0.291 $\pm$ 0.032   & 0.170 $\pm$ 0.036   &      0.340:\\
 4889.70 & [Fe II] &  3F & -0.007 &      0.028:    &        \nodata      &        \nodata      &        \nodata\\
 4921.93 &  He I &  48 & -0.015 & 1.047 $\pm$ 0.068 & 1.016 $\pm$ 0.057   & 0.980 $\pm$ 0.069   &  0.82 $\pm$ 0.19\\
 4931.32 &  [O III] & 1F & -0.017 & 0.083 $\pm$ 0.027   & 0.056 $\pm$ 0.017   &     0.030:   &      \nodata\\
 4958.91 &  [O III] & 1F & -0.024 & 204.0 $\pm$ 6.1     & 206.2 $\pm$ 6.2  & 160.9 $\pm$ 4.8     & 104.3 $\pm$ 3.3\\
 4985.90 & [Fe III] & 2F & -0.031 & 0.436 $\pm$ 0.061   & 0.459 $\pm$ 0.055  & 0.503 $\pm$ 0.051 &  1.85 $\pm$ 0.29\\
 5006.84 &  [O III] & 1F & -0.036 &   579 $\pm$ 17    &   597 $\pm$ 18      &   460 $\pm$ 13      & 300.2 $\pm$ 9.3\\
 5015.68 &  He I & 4F & -0.038 &  1.95 $\pm$ 0.12    &  2.18 $\pm$ 0.12    &  1.98 $\pm$ 0.11    &  1.76 $\pm$ 0.29\\
 5041.03 &  Si II & 5 & -0.044 & 0.144 $\pm$ 0.039   & 0.205 $\pm$ 0.039   & 0.124 $\pm$ 0.026   &  0.53 $\pm$ 0.18\\
 5047.74 &  He I & 47 & -0.046 & 0.277 $\pm$ 0.051   & 0.327 $\pm$ 0.048   & 0.311 $\pm$ 0.040   &  2.24 $\pm$ 0.32\\
 5055.98 &  Si II &  5 & -0.048 & 0.346 $\pm$ 0.055   & 0.223 $\pm$ 0.040   & 0.300 $\pm$ 0.039   &        \nodata\\
 5158.81 & [Fe II] &  19F & -0.073 &        \nodata      & 0.118 $\pm$ 0.031   & 0.181 $\pm$ 0.031   &   0.382:\\
 5191.82 & [Ar III] &  3F & -0.081 & 0.158 $\pm$ 0.040   & 0.153 $\pm$ 0.034   & 0.089 $\pm$ 0.022   &   0.297:\\
 5197.90 & [N I] &  1F & -0.082 & 0.318 $\pm$ 0.053  & 0.301 $\pm$ 0.045   & 0.220 $\pm$ 0.034   &  0.76 $\pm$ 0.20\\
 5200.26 & [N I] &  1F & -0.083 & 0.251 $\pm$ 0.048   & 0.237 $\pm$ 0.041   & 0.147 $\pm$ 0.028   &        \nodata\\
 5261.61 & [Fe II] & 19F & -0.098 &      0.074:   &      0.042:    & 0.061 $\pm$ 0.019   &        \nodata\\
 5270.40 & [Fe III] & 1F & -0.100 & 0.466 $\pm$ 0.061  & 0.445 $\pm$ 0.053   & 0.324 $\pm$ 0.040 &  0.77 $\pm$ 0.21\\
 5517.71 & [Cl III] & 1F & -0.154 & 0.375 $\pm$ 0.055   & 0.320 $\pm$ 0.045   & 0.362 $\pm$ 0.042   &      0.254:\\
 5537.88 & [Cl III] & 1F & -0.158 & 0.289 $\pm$ 0.050   & 0.251 $\pm$ 0.040   & 0.271 $\pm$ 0.037   &      0.105:\\
 5754.64 &  [N II] & 3F & -0.194 & 0.500 $\pm$ 0.062   & 0.439 $\pm$ 0.050   & 0.177 $\pm$ 0.030   &      0.316:\\
 5875.64 &  He I & 11 & -0.215 & 11.50 $\pm$ 0.43    & 12.34 $\pm$0.45     & 11.14 $\pm$ 0.42   & 10.74 $\pm$ 0.65\\
 5957.56 &  Si II &  4 & -0.228 & 0.160 $\pm$ 0.039   &        \nodata   &        \nodata   &        \nodata\\
 5978.93 &  Si II &  4 & -0.231 &        \nodata   &      0.035:            & 0.066 $\pm$ 0.019   &        \nodata\\
 6300.30 & [O I] & 1F & -0.282 &  2.32 $\pm$ 0.13    &  2.15 $\pm$ 0.12   &  2.35 $\pm$ 0.12   &  6.05 $\pm$ 0.48\\
 6312.10 & [S III] & 3F & -0.283 &  2.51 $\pm$ 0.14  &  2.43 $\pm$ 0.13   & 1.705 $\pm$ 0.099   &  1.53 $\pm$ 0.26\\
 6347.11 &  Si II &  2 & -0.289 & 0.119 $\pm$ 0.034  & 0.076 $\pm$ 0.023   & 0.108 $\pm$ 0.023   &    \nodata\\
 6363.78 &  [O I] & 1F & -0.291 & 0.730 $\pm$ 0.072  & 0.677 $\pm$ 0.060   & 0.739 $\pm$ 0.060   &  1.90 $\pm$ 0.29\\
 6371.36 &  Si   II & 2 & -0.292 & 0.069 $\pm$ 0.027   & 0.130 $\pm$ 0.029   & 0.116 $\pm$ 0.024   &      0.137:\\
 6548.03 &  [N II] & 1F & -0.318 &  9.23 $\pm$ 0.39   &  8.21 $\pm$ 0.34   &  3.89 $\pm$ 0.18   &  7.63 $\pm$ 0.54\\
 6562.82 &   H I &   \Ha & -0.320 &   282 $\pm$ 10      &   284 $\pm$ 10   &   288 $\pm$ 10   & 285.1 $\pm$ 9.8\\
 6578.05 &   C II &    2 & -0.322 &        \nodata   &        \nodata       & 0.071 $\pm$ 0.019   &        \nodata\\
 6583.41 &  [N II] &  1F & -0.323 &  29.2 $\pm$ 1.1    & 24.61 $\pm$ 0.95   & 11.83 $\pm$ 0.48   &  22.5 $\pm$ 1.1\\
 6678.15 &  He I &  46 & -0.336 &  3.38 $\pm$ 0.17   &  3.32 $\pm$ 0.16   &  3.00 $\pm$ 0.15   &  2.83 $\pm$ 0.34\\
 6716.47 &  [S II] &  2F & -0.342 & 13.37 $\pm$ 0.55   & 11.52 $\pm$ 0.47   & 13.90 $\pm$ 0.57   &  36.0 $\pm$ 1.5\\
 6730.85 &  [S II] &  2F & -0.344 & 12.44 $\pm$ 0.51   & 11.03 $\pm$ 0.45   & 12.17 $\pm$ 0.50   &  29.0 $\pm$ 1.3\\
 7002.23 &   O I &    21 & -0.379 &      0.071:        & 0.087 $\pm$ 0.022   & 0.085 $\pm$ 0.020   &      0.244:\\
 7065.28 &  He I &  1/10 & -0.387 &  4.36 $\pm$ 0.20   &  5.35 $\pm$ 0.23   &  3.04 $\pm$ 0.14   &  2.19 $\pm$ 0.23\\
 7135.78 & [Ar III] & 1F & -0.396 & 13.51 $\pm$ 0.56   & 13.07 $\pm$ 0.54   & 10.40 $\pm$ 0.44   &  8.89 $\pm$ 0.45\\
 7155.14 & [Fe II] &   14F & -0.399 &      0.047:        & 0.053 $\pm$ 0.019   & 0.048 $\pm$ 0.016   &      0.094:\\
 7160.13 &  He I &  1/10 & -0.399 &        \nodata   &      0.015:   &        \nodata   &        \nodata\\
 7281.35 &  He I &    45 & -0.414 & 0.579 $\pm$ 0.068   &        \nodata   & 0.451 $\pm$ 0.040   &        \nodata\\
 7318.39 &  [O II] &  2F & -0.418 &  2.73 $\pm$ 0.15   &  2.74 $\pm$ 0.13   &  2.78 $\pm$ 0.13   &  4.86 $\pm$ 0.32\\
 7329.66 &  [O II] &  2F & -0.420 &  2.09 $\pm$ 0.12   &  2.10 $\pm$ 0.10   &  2.23 $\pm$ 0.11   &  3.56 $\pm$ 0.28\\
 7377.83 & [Ni II] &  2F & -0.425 &      0.074:        & 0.052 $\pm$ 0.019   & 0.083 $\pm$ 0.020   &      0.312:\\
 7411.61 & [Ni II] &   2F & -0.429 &      0.019:   &      0.011:   &      0.016:   &        \nodata\\
 7423.64 &   N I &  3 & -0.431 &      0.011:   &      0.010:   &      0.007:   &        \nodata\\
 7442.30 &   N I &  3 & -0.433 &      0.028:   &      0.032:   &      0.035:   &      0.098:\\
 7452.50 & [Fe II] &   14F & -0.434 &      0.024:   &      0.018:   &      0.017:   &        \nodata\\
 7468.31 &   N I &     3 & -0.436 &      0.055:        & 0.061 $\pm$ 0.020   & 0.071 $\pm$ 0.019   &      0.159:\\
 7499.85 &  He I &   1/8 & -0.439 &      0.029:   &      0.025:              & 0.360 $\pm$ 0.036   &        \nodata\\
 7530.60 &   C II & 16.08 & -0.443 &      0.049:        & 0.051 $\pm$ 0.018   &      0.023:   &        \nodata\\
 7751.10 & [Ar III] & 2F & -0.467 &  3.48 $\pm$ 0.18   &  3.15 $\pm$ 0.15   &  2.64 $\pm$ 0.13   &  2.17 $\pm$ 0.23\\
 8000.08 & [Cr III] & 1F & -0.492 &      0.037:        & 0.047 $\pm$ 0.017   &      0.037:   &        \nodata\\
 8045.63 & [Cl IV] & 1F & -0.497 & 0.109 $\pm$ 0.042  & 0.141 $\pm$ 0.025   &      0.032:   &        \nodata\\
 8084.00 &  He I &  4/18 & -0.500 &      0.028:   &        \nodata   &        \nodata   &        \nodata\\
 8125.31 &  Ca I] &    :: & -0.504 &      0.012:   &        \nodata   &        \nodata   &        \nodata\\
 8210.72 &   N I &     2 & -0.512 &      0.014:   &      0.024:   &        \nodata   &        \nodata\\
 8216.34 &   N I &     2 & -0.513 &      0.036:   &      0.026:   &        \nodata   &        \nodata\\
 8223.14 &   N I &     2 & -0.514 &      0.046:   & 0.041 $\pm$ 0.016   &        \nodata   &        \nodata\\
 8271.93 &   H I &   P33 & -0.518 &        \nodata   &        \nodata   & 0.045 $\pm$ 0.015   &        \nodata\\
 8276.31 &   H I &   P32 & -0.518 &        \nodata   &        \nodata   & 0.046 $\pm$ 0.016   &        \nodata\\
 8281.12 &   H I &   P31 & -0.519 &        \nodata   &        \nodata   & 0.059 $\pm$ 0.017   &        \nodata\\
 8286.43 &   H I &   P30 & -0.519 &        \nodata   &        \nodata   & 0.091 $\pm$ 0.020   &        \nodata\\
 8298.83 &   H I &   P28 & -0.521 &        \nodata   & 0.069 $\pm$ 0.018   & 0.084 $\pm$ 0.019   &        \nodata\\
 8306.11 &   H I &   P27 & -0.521 &        \nodata   & 0.046 $\pm$ 0.015   & 0.066 $\pm$ 0.018   &        \nodata\\
 8314.26 &   H I &   P26 & -0.522 &      0.089:   & 0.113 $\pm$ 0.022   & 0.095 $\pm$ 0.020   &        \nodata\\
 8323.42 &   H I &   P25 & -0.523 &      0.095:   & 0.104 $\pm$ 0.021   & 0.143 $\pm$ 0.024   &        \nodata\\
 8333.78 &   H I &   P24 & -0.524 &        \nodata   & 0.692 $\pm$ 0.050   & 0.903 $\pm$ 0.061   &        \nodata\\
 8345.55 &   H I &   P23 & -0.525 & 0.111 $\pm$ 0.042   & 0.105 $\pm$ 0.021   & 0.156 $\pm$ 0.025   &    \nodata\\
 8359.00 &   H I &   P22 & -0.526 & 0.181 $\pm$ 0.047   & 0.180 $\pm$ 0.026   & 0.217 $\pm$ 0.028   &    \nodata\\
 8361.67 &  He I &   1/6 & -0.526 &   0.073:   & 0.059 $\pm$ 0.018    & 0.060 $\pm$ 0.017   &    \nodata\\
 8374.48 &   H I &   P21 & -0.527 & 0.190 $\pm$ 0.048  & 0.182 $\pm$ 0.026  & 0.214 $\pm$ 0.028  &     \nodata\\
 8392.40 &   H I &   P20 & -0.529 & 0.260 $\pm$ 0.052  & 0.232 $\pm$ 0.029  & 0.264 $\pm$ 0.031  &  0.49 $\pm$ 0.14\\
 8413.32 &   H I &   P19 & -0.531 & 0.274 $\pm$ 0.052  & 0.229 $\pm$ 0.029   & 0.290 $\pm$ 0.032  &  31.6 $\pm$ 1.3\\
 8437.96 &   H I &   P18 & -0.533 & 0.319 $\pm$ 0.055   & 0.300 $\pm$ 0.032   & 0.320 $\pm$ 0.034  &   0.137:\\
 8446.36 &   O I &   4 & -0.534 & 0.645 $\pm$ 0.069  & 0.676 $\pm$ 0.049   & 0.688 $\pm$ 0.051   &  0.76 $\pm$ 0.16\\
 8467.25 &   H I &   P17 & -0.536 & 0.373 $\pm$ 0.057 & 0.375 $\pm$ 0.036   & 0.369 $\pm$ 0.036 &  0.35 $\pm$ 0.13\\
 8486.27 &  He I &  6/16 & -0.537 &      0.019:   &      0.018:   &        \nodata   &        \nodata\\
 8502.48 &   H I &   P16 & -0.539 & 0.446 $\pm$ 0.061  & 0.454 $\pm$ 0.039  & 0.463 $\pm$ 0.041 &  0.43 $\pm$ 0.14\\
 8665.02 &   H I &   P13 & -0.553 & 0.796 $\pm$ 0.075   & 1.003 $\pm$ 0.064   & 0.827 $\pm$ 0.058 & 0.59 $\pm$ 0.15\\
 8680.28 &   N I &     1 & -0.554 &     0.049:         & 0.043 $\pm$ 0.016   & 0.044 $\pm$ 0.015   &        \nodata\\
 8703.25 &   N I &     1 & -0.556 &       \nodata   &      0.024:   &      0.016:   &        \nodata\\
 8711.70 &   N I &     1 & -0.556 &     0.027:         & 0.059 $\pm$ 0.018   & 0.054 $\pm$ 0.016   &        \nodata\\
 8718.83 &   N I &     1 & -0.557 &       \nodata   &      0.020:   &        \nodata   &        \nodata\\
 8733.43 &  He I &  6/12 & -0.558 &      0.019:   &      0.024:   &      0.013:   &        \nodata\\
 8750.47 &   H I &   P12 & -0.560 & 1.057 $\pm$ 0.086   & 0.829 $\pm$ 0.056  & 1.442 $\pm$ 0.087 & 1.02 $\pm$ 0.17\\
 8845.38 &  He I &  6/11 & -0.567 &      0.010:   &      0.027:   &      0.020:   &        \nodata\\
 8862.79 &   H I &   P11 & -0.569 &  1.37 $\pm$ 0.10 & 1.400 $\pm$ 0.082   & 1.430 $\pm$ 0.087   &  1.32 $\pm$ 0.19\\
 9014.91 &   H I &   P10 & -0.581 &  1.70 $\pm$ 0.11 & 1.551 $\pm$ 0.090   &  2.22 $\pm$ 0.13   &  1.67 $\pm$ 0.21\\
 9068.90 &  [S III] &    1F & -0.585 &  25.0 $\pm$ 1.3  &  25.5 $\pm$ 1.3   &  21.6 $\pm$ 1.1   & 17.94 $\pm$ 0.82\\
 9123.60 & [Cl II] &    1F & -0.589 &      0.039:   &      0.036:   &        \nodata   &        \nodata\\
 9229.01 &   H I &    P9 & -0.596 &  2.54 $\pm$ 0.15   &  2.63 $\pm$ 0.14   &  2.64 $\pm$ 0.15   &  2.65 $\pm$ 0.25\\
 9530.60 &  [S III] &  1F & -0.618 &  71.0 $\pm$ 8.5   &  63.8 $\pm$ 7.0     &  51.3 $\pm$ 6.2   &  43.7 $\pm$ 5.2\\
 9545.97 &   H I &    P8 & -0.619 &  2.95 $\pm$ 0.18   &  3.20 $\pm$ 0.18   &  4.67 $\pm$ 0.25   &  2.69 $\pm$ 0.25\\
10031.20 &  He I &   7/7 & -0.649 & 0.214 $\pm$ 0.047   &        \nodata    &        \nodata   &        \nodata\\
10049.37 &   H I &    P7 & -0.650 &  6.18 $\pm$ 0.35   &  6.65 $\pm$ 0.36   &  5.62 $\pm$ 0.31   & 11.73 $\pm$ 0.60\\
\hline
\enddata
\end{deluxetable}

\clearpage

The observed line intensities must be corrected for interstellar reddening. This can be done using the reddening 
constant, $c(H\beta)$, obtained from the intensities of \ion{H}{1} lines. However, the fluxes of \ion{H}{1} lines are 
also affected by underlying stellar absorption. Consequently, we have performed an iterative procedure to derive 
both $c(H\beta)$ and the equivalent widths of the absorption in the hydrogen lines, $W_{abs}$, that we use 
to correct the observed line intensities. As we have both \ion{H}{1} Balmer and Paschen lines, we perform this 
procedure separately for them, following the procedure explained in \citet*{LSEGR06}.

\begin{deluxetable}{lcccc} 
\tabletypesize{\footnotesize}
\tablecaption{H$\beta$ fluxes, reddening coefficients and equivalent widths.
\label{reddening}}
\tablewidth{0pt}
\tablehead{
\colhead{} &
\colhead{(A) HII-2} &
\colhead{(B) HII-1} & 
\colhead{(C) UV-1} &
\colhead{(D) UV-2}} 
\startdata

    \noalign{\smallskip}
    $F$(H$\beta$) ($\times 10^{14}$ erg s$^{-1}$ cm$^{-2}$) .................. &
	          13.45 $\pm$ 0.43   &  13.52 $\pm$ 0.42  &  10.13 $\pm$ 0.33 & 2.56 $\pm$ 0.09 \\ 
   \noalign{\smallskip}
   $W$(H$\alpha$) (\AA) .................. &
              919    &  1009    &  470   &   169   \\
   $W$(H$\beta$) (\AA) .................. &  
              234     &  254   &  94  &   39  \\
   $W$(H$\gamma$) (\AA) .................. &    
              96      &  93    &  43  &   44  \\
   $W$(H$\delta$) (\AA) .................. &
              49      &  45    &  18  &   23 \\
   $W$(H$\epsilon$) (\AA) .................. &
              27      &  25    &  11  &   5  \\
   $W$(H9) (\AA) .................. &
	          12      &  10    &   4  &   2 \\
	\noalign{\smallskip} 
	$W_{abs}$ (\Ha) (\AA) &   0.0 $\pm$ 0.2  &  0.0 $\pm$ 0.2  & 0.0 $\pm$ 0.2  & 0.7 $\pm$ 0.2  \\       
	$W_{abs}$ (\Hd) (\AA) &   0.7 $\pm$ 0.2  &  1.2 $\pm$ 0.2  & 0.4 $\pm$ 0.2  & 0.8 $\pm$ 0.2  \\
	$W_{abs}$ (\Hg) (\AA) &   2.2 $\pm$ 0.2  &  3.9 $\pm$ 0.2  & 1.1 $\pm$ 0.2  & 0.8 $\pm$ 0.2  \\
	$W_{abs}$ (\He) (\AA) &   1.2 $\pm$ 0.2  &  1.5 $\pm$ 0.2  & 0.7 $\pm$ 0.2  & 0.4 $\pm$ 0.3  \\ 
	$W_{abs}$ (H9) (\AA)  &   0.4 $\pm$ 0.3  &  0.9 $\pm$ 0.3  & 0.2 $\pm$ 0.3  & 0.2 $\pm$ 0.3  \\
	
	\noalign{\smallskip} 
   $W_{abs}$ (Balmer) (\AA) &
              1.3 $\pm$ 0.3 &    1.7 $\pm$ 0.3  & 0.8 $\pm$ 0.2    &  0.6 $\pm$ 0.3 \\
   $c$(H$\beta$) (Balmer)    &
             0.22 $\pm$ 0.02 &  0.36 $\pm$ 0.03 & 0.23 $\pm$ 0.03 & 0.09 $\pm$ 0.02 \\  
    \noalign{\smallskip}
   $W_{abs}$ (Paschen) (\AA)  &
              0.0 $\pm$ 0.1 &    0.0 $\pm$ 0.1 &   0.0 $\pm$ 0.1 &  0.0 $\pm$ 0.1 \\
    $c$(H$\beta$) (Paschen)    &
             0.24 $\pm$ 0.02 &  0.39 $\pm$ 0.02 & 0.27 $\pm$ 0.02 & 0.10 $\pm$ 0.02 \\
    \noalign{\smallskip}
	\noalign{\smallskip}
   $c$(H$\beta$) (adopted)    &
       {\bf  0.23 $\pm$ 0.02} & {\bf  0.38 $\pm$ 0.03} & {\bf  0.25 $\pm$ 0.03} & {\bf  0.10 $\pm$ 0.02}\\
	\noalign{\smallskip}
	\noalign{\smallskip}    
\hline
\enddata
\end{deluxetable}

\begin{figure}[t]
\includegraphics[angle=270,width=1\linewidth]{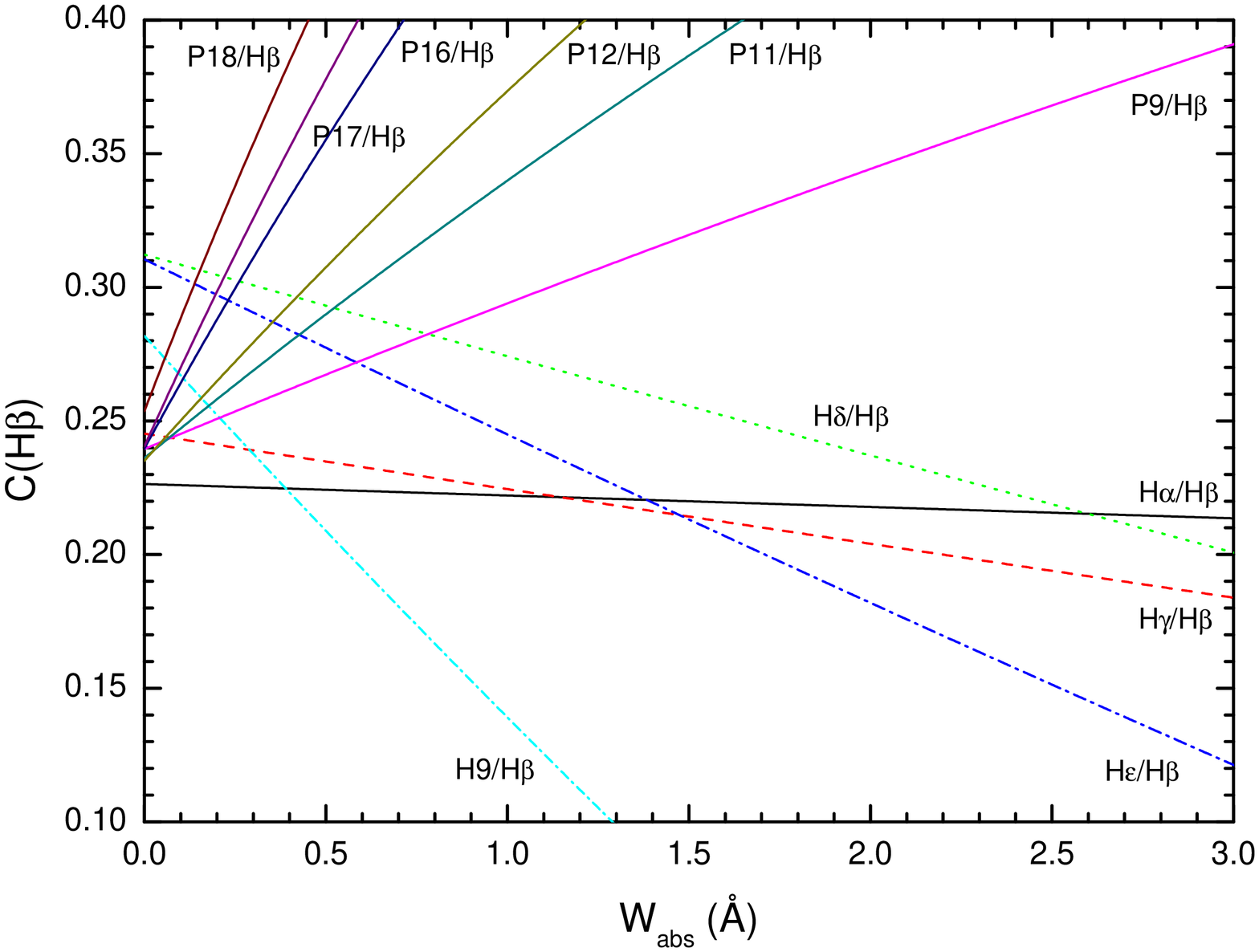}
\caption{\small{Reddening constant, $c(H\beta)$, as a function of the absorption equivalent width in the hydrogen 
lines. Each straight line represent the behaviour of the ratio of a given Balmer and Paschen line with respect to 
H$\beta$.}}
\label{chb}
\end{figure}

In Figure~\ref{chb} we plot the reddening constant, $c(H\beta)$, versus the absorption equivalent widths in the 
hydrogen lines, $W_{abs}$, for the data of region A. It can be noted that absorption in the \ion{H}{1} Paschen lines 
is practically negligible. This result is also found in the rest of analyzed regions. However, \ion{H}{1} Balmer 
lines have considerably differences in their $W_{abs}$, although the $c(H\beta)$ derived from them is very similar to 
those  derived from Paschen lines. We have assumed the average value between the $c(H\beta)$ derived from Balmer and 
Paschen lines as representative of the region. This value can be used to estimate the appropriate value of $W_{abs}$ 
for each Balmer line. The $c(H\beta)$ and $W_{abs}$ that provide the best match between the corrected and the 
theoretical Balmer and Paschen line ratios, as well as the finally adopted values of $c(H\beta)$ and $W_{abs}$ for the 
brightest Balmer lines are shown in Table~\ref{reddening}.

\section{Emission line profiles \\ and gas kinematics}

Our observations confirm the presence of asymmetric wings in the line profiles at the center of NGC 5253, as were previously reported in echelle 
spectra by \citet{Martin95} and \citet{Marlowe95}. The line-profiles of regions A and B show a broader intense component underlying a narrower one 
with velocity shifts of the order of lower than 10 km s$^{-1}$. This broad 
component is redshifted with respect to the narrow one in zone A, but appears slightly blueshifted in region B. Region C seems to show a rather faint 
broad component. Finally, the line profiles of 
region D seem to be composed by the combination of two similar narrow components with a velocity shift of $\sim$ 35 km s$^{-1}$. Consistent values of 
the velocity shift between components were found by \citet{Martin95} and \citet{Marlowe95}. 

The velocity separation between components reaches a maximum at zone D. The maximum width of the broad 
component occurs in regions A and B where the full width at half 
maximum (FWHM) reaches values of the order of 110 km s$^{-1}$. \citet{Martin95} report consistent values of the 
FWHM of this component. The velocity shifts and FWHM ratio between broad and narrow component for other lines like [\ion{O}{2}], [\ion{O}{3}], 
[\ion{N}{2}], and [\ion{S}{2}] are qualitatively similar for a given zone. Similar complex emission-line profiles and FWHM of the broad components 
were found by \citet{Marlowe95} and M\'endez \& Esteban (1997) for different samples of dwarf starburst galaxies.



\begin{figure}[t!]
\includegraphics[angle=270,width=1\linewidth]{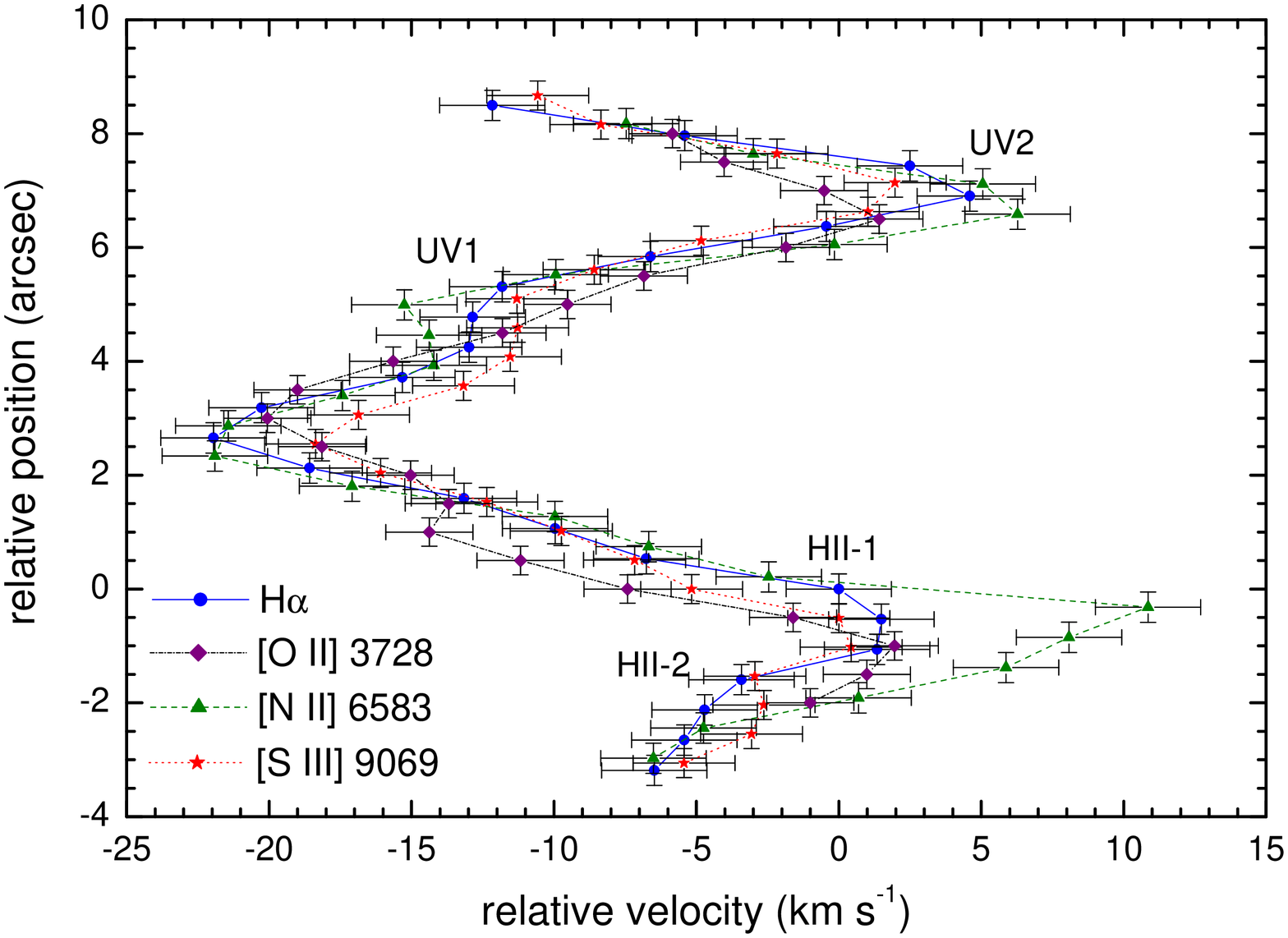}
\caption{\small{Position-velocity diagram for the slit position observed using \Ha, [O II] $\lambda\lambda$3726,3729, [N II] 
$\lambda$6583 and [S III] $\lambda$9069 emission lines. The abscissa shows velocities with respect to the center of zone B.}}
\label{pv}
\end{figure}

The high signal-to-noise of our spectra has permitted to perform a spatially resolved  analysis of the kinematics of the ionized gas observed in 
the center of NGC 5253. It was studied via the analysis of the centroids of the total line profiles of \Ha, 
[\ion{O}{2}] $\lambda\lambda$3726,3729, [\ion{N}{2}] $\lambda$6583 and [\ion{S}{3}] $\lambda$9069 
along the slit position. We have extracted zones of 3 pixels wide that corresponds to 0.54$\arcsec$ in the sky for the \Ha\ and [\ion{N}{2}] emission 
lines 
and 0.51$\arcsec$ for the [\ion{S}{3}] line. The width of 
the extracted zones for the position of [\ion{O}{2}] is 2 pixels, that corresponds to 0.50$\arcsec$. In Fi\-gure~\ref{pv} we show the 
position-velocity (PV) diagrams obtained. All the velocities are referred to the velocity of the center of region B 
(HII-1). 

Figure~\ref{pv} indicates that the ionized gas at the center of NGC 5253 follows a sinusoidal pattern, indicating 
that its kinematics is not due to pure rotation. The behaviour of our PV diagram is in agreement (considering the shape and the velocity amplitudes) 
with that obtained by \citet{Martin95} for their slit positions 2 and 3 in the central zones of the galaxy. 
The sinusoidal or wave-like pattern may be produced by distortions produced by the presence of dynamically decoupled gas systems (e.g. Schweizer 
1982).  This explanation would 
imply that a merging process is ongoing in the central zone of the galaxy. Another possibility is that the velocity pattern is produced by the 
outflow from the central starbursts. In fact, the wave-like forms produced by supershells are rather common in starbursting dwarf galaxies 
\citep{Martin95,Marlowe95,Pustilnik04}. We also find that the PV diagrams of the narrow and broad components are similar, it would imply 
that the outflows affect the dynamics of both components in a similar way. 
It is remarkable that the PV diagrams derived from emission lines of different ions are in excellent agreement {\it except} for the [\ion{N}{2}] 
$\lambda$6583 line at regions A (HII-2) and B (HII-1). This fact could be related to a possible localized nitrogen pollution 
in those zones (we discuss this in \S 12).

\begin{deluxetable}{lccccc}
\tabletypesize{\scriptsize}
\tablecaption{Physical Conditions of the Ionized Gas\label{temden}}
\tablewidth{0pt}
\tablehead{
\colhead{Diagnostic} & 
\colhead{Lines}  & 
\colhead{(A) HII-2} &
\colhead{(B) HII-1} &
\colhead{(C) UV-1} &
\colhead{(D) UV-2}}
\startdata
$n_{\rm e}$ (cm$^{-3}$)& [O\thinspace II] ($\lambda$3726)/($\lambda$3729)& 
                660$^{+130}_{-140}$   & 600$^{+120}_{-130}$ & 420$\pm 100$ & 270 $\pm$ 80 \\
& [S\thinspace II] ($\lambda$6716)/($\lambda$6731)& 
                460$^{+160}_{-210}$   & 530$^{+170}_{-220}$ & 330$^{+130}_{-180}$ & 190$^{+110}_{-150}$ \\
& [N\thinspace I]  ($\lambda$5198)/($\lambda$5200)&
                670 $^{+550}_{<100}$  & 670$^{+510}_{<100}$ & 1140:: & \nodata \\  
& [Cl\thinspace III] ($\lambda$5518)/($\lambda$5538)& 
                530:                  & 640:                & 380:   & \nodata \\   
& [Fe III] & 
                750 $\pm$ 250 & 650 $\pm$ 350 & 350 $\pm$ 300  & 300 $\pm$ 300  \\
& [Ar IV] ($\lambda$4711)/($\lambda$4740)& 
                5600$^{+3400}_{-2600}$& 5100$^{+2200}_{-1900}$& 1300$^{+7200}_{-1300}$ & \nodata	  \\
& & \\
& {\bf Adopted value}$^{\rm b}$ & {\bf 580 $\pm$ 110} & {\bf 610 $\pm$ 100 }& {\bf 370 $\pm$ 80} &{\bf 230 $\pm$ 70 }\\

& & \\

$T_{\rm e}$ (K) High & [O\thinspace III] ($\lambda$4959+$\lambda$5007)/($\lambda$4363) & 
                11960$^{+270}_{-300}$  & 12010$^{+270}_{-300}$ & 10940$^{+240}_{-270}$ & 10990$^{+490}_{-620}$\\
& [S\thinspace III] ($\lambda$9069+$\lambda$9532)/($\lambda$6312)$^{\rm a}$ & 
				12330$^{+640}_{-800}$ & 11970$^{+580}_{-720}$& 10910$^{+510}_{-630}$& 11300$^{+1000}_{-1600}$\\
& [Ar\thinspace III] ($\lambda$7136+$\lambda$7751)/($\lambda$5192)& 
                12000$^{+1200}_{-2200}$& 12100$^{+1100}_{-1900}$& 10600$^{+940}_{-1600}$& \nodata \\
& & \\
&{\bf Adopted value}&{\bf 12100 $\pm$ 260} & {\bf 12030 $\pm$ 260 }& {\bf 10810 $\pm$ 230}&{\bf 11160 $\pm$ 510 }\\

& & \\

$T_{\rm e}$ (K) Low & [O\thinspace II] ($\lambda$3726+$\lambda$3729)/($\lambda$7320+$\lambda$7330)& 
                11300$^{+600}_{-740}$  & 11330$^{+560}_{-680}$ & 10660$^{+490}_{-590}$ & 10570$^{+580}_{-740}$\\
& [S\thinspace II] ($\lambda$6716+$\lambda$6731)/($\lambda$4069+$\lambda$4076)&  
				10880$^{+860}_{-1160}$ &  8680$^{+550}_{-720}$ &  8180$^{+530}_{-710}$ &  8380$^{+760}_{-1130}$\\
& [N\thinspace II] ($\lambda$6548+$\lambda$6583)/($\lambda$5755) & 
                11040$^{+700}_{-960}$ & 11170$^{+680}_{-920}$ & 10410$^{+780}_{-11600}$ & 10100$^{+1500}_{-4200}$ \\
& & \\
& {\bf Adopted value}$^{\rm c}$ & {\bf 11170 $\pm$ 520} & {\bf 11250 $\pm$ 490}&{\bf 10530 $\pm$ 470}&{\bf 10350 $\pm$ 650}\\
& & \\
$T_{\rm e}$ (K)& He\thinspace I  & 10270 $\pm$ 300 & 10200 $\pm$ 300 &  9000 $\pm$ 300& 10600 $\pm$ 450 \\
\hline
\enddata
\tablenotetext{a}{[S\thinspace III] $\lambda$9532 affected by a sky emission line, so we used the 
theoretical ratio $\lambda$9532/$\lambda$9069 = 2.48.}
\tablenotetext{b}{$n_e$(Ar\thinspace IV) was not considered in the average, see text.}
\tablenotetext{c}{$T_e$(S\thinspace II) was not considered in the average, see text.}
\end{deluxetable}

\section{Physical conditions of the ionized gas}

We have derived the electron temperature, $T_e$, and density, $n_e$, of the ionized gas using several emission line 
ratios. The values obtained for each region are compiled in Table~\ref{temden}. All determinations were computed with 
the IRAF task TEMDEN on the NEBULAR package \citep{SD95}, except for the density derived from [\ion{Fe}{3}] emission 
lines (see below). We have changed the default atomic data of O$^+$, S$^+$, and S$^{++}$ included in the last version 
of NEBULAR (February 2004) by other datasets that we consider give better results. These changes are indicated in 
Table 4 of \citet{GRE04}.

In order to derive $n_e$, we have used several emission line ratios between CELs (see Table~\ref{temden}) and the intensities of several 
[\ion{Fe}{3}] lines. The [\ion{Fe}{3}] density was computed 
from the intensity of the brightest lines of each region (those with errors less than 30\% and not affected by line 
blending) and applying the computations of \citet{Ro02}. All the values of $n_e$ are consistent within the errors (see 
Table~\ref{temden}) except those obtained from the [\ion{Ar}{4}] lines, which are always higher and are expected to be representative of the inner 
zones of the nebulae. In the case of zones A and B that difference is almost one order of magnitude indicating a strong density 
stratification in the ionized gas.

Once $n_e$ was obtained we used it to derive $T_e$ using [\ion{O}{3}], [\ion{S}{3}], [\ion{Ar}{3}], 
[\ion{O}{2}], [\ion{S}{2}], and [\ion{N}{2}] line ratios, and we iterated until convergence. We have determined the 
characteristic temperature of the helium zone, $T_e$(\ion{He}{1}), in the presence of temperature fluctuations (see 
\S 7) using the formulation of \citet*{PPL02}. All determinations of $T_e$ are included in Table~\ref{temden}. It is important to remark 
that previous $T_e$ determinations in the literature were restricted to a single indicator: $T_e$(\ion{O}{3}). The values of $T_e$(\ion{O}{3}) 
determined by \citet{Kobulnicky97} are consistent with 
those derived from our data. 

We have assumed a two-zone approximation to describe the temperature structure of the nebulae. Therefore, we used the 
mean of $T_e$(\ion{O}{3}), $T_e$(\ion{S}{3}) and $T_e$(\ion{Ar}{3}) as the representative temperature for high 
ionization potential ions whereas the mean of $T_e$(\ion{O}{2}) and $T_e$(\ion{N}{2}) was assumed for the low 
ionization potential ion temperatures. We do not include $T_e$(\ion{S}{2}) in the average because its value is around 
1500 K lower that the ones obtained from [\ion{O}{2}] and [\ion{N}{2}] ratios in three of the zones. This difference 
is also reported in the Galactic \ion{H}{2} region S 311 by \citet{GRE05} and might be produced by the 
presence of a temperature stratification in the outer zones of the nebulae. 

\section{Ionic Abundances}

The large collection of emission lines measured in the spectra has permitted the derivation of abundances for an 
unprecedented large number of ions in NGC 5253 from both recombination and collisionally excited 
lines. 

\subsection{He$^+$ abundance}

\begin{deluxetable}{c@{\hspace{10pt}}c@{\hspace{10pt}}c@{\hspace{10pt}}c@{\hspace{10pt}}c@{\hspace{10pt}}}
\tabletypesize{\scriptsize}
\tablecaption{He$^{+}$/H$^+$ ratios$^{\rm a}$
\label{helio}}
\tablewidth{0pt}
\tablehead{
\colhead{Line} & 
\colhead{(A) HII-2} & 
\colhead{(B) HII-1} &
\colhead{(C) UV-1} &
\colhead{(D) UV-2}}
\startdata
3819.61 	& 830 $\pm$ 166		& 704 $\pm$ 152	& 707 $\pm$ 106 & \nodata 	\\
3964.73 	& 642 $\pm$ 88 		& 603 $\pm$ 46 	& 579 $\pm$ 57 	& \nodata 	\\ 
4026.21 	& 821 $\pm$ 43 		& 775 $\pm$ 38 	& 772 $\pm$ 42 	& 653 $\pm$ 108 \\
4387.93 	& 676 $\pm$ 79 		& 769 $\pm$ 69 	& 649 $\pm$ 81 	& \nodata 	\\
4471.09 	& 876 $\pm$ 34 		& 794 $\pm$ 29 	& 736 $\pm$ 27 	& 676 $\pm$ 68 	\\
4713.14 	& 899 $\pm$ 84 		& 826 $\pm$ 61 	& 691 $\pm$ 84 	& \nodata 	\\
4921.93 	& 775 $\pm$ 50 		& 759 $\pm$ 43 	& 722 $\pm$ 51 	& 625 $\pm$ 145 \\
5875.64 	& 777 $\pm$ 29 		& 840 $\pm$ 31 	& 758 $\pm$ 29 	& 786 $\pm$ 48 	\\
6678.15 	& 836 $\pm$ 42 		& 838 $\pm$ 41 	& 735 $\pm$ 37 	& 741 $\pm$ 90 	\\
7065.28 	& 803 $\pm$ 37 		& 786 $\pm$ 34 	& 729 $\pm$ 34 	& 732 $\pm$ 77 	\\
7281.35 	& 775 $\pm$ 91 		& \nodata	& 632 $\pm$ 56 	& \nodata 	\\
Adopted$^{\rm b}$& 807 $\pm$ 16 	& 791 $\pm$ 14 	& 729 $\pm$ 13 	& 737 $\pm$ 34 \\
$\tau_{3889}$ 	& 9.61 $\pm$ 0.74  	& 11.68 $\pm$ 0.70& 7.33 $\pm$ 0.70& 1.61 $\pm$ 0.92 \\
$\chi^2$ 	& 18.19 		& 19.54 	& 12.67 	& 3.35 \\
\enddata
\tablenotetext{a}{In units of 10$^{-4}$}
\tablenotetext{b}{The adopted value includes all the relevant uncertainties in emission line intensities, $n_e$, and $\tau_{3889}$, and 
assuming the adopted $t^2$ values.}
\end{deluxetable}

We have measured a large number of \ion{He}{1} lines in our spectra of NGC 5253, the maximum in region A where we 
detect 34 lines. This permits to determine the He$^+$/H$^+$ ratio with much better precision than in previous studies. We have followed the same 
method explained in \citet{GRE05}, who applied the maximum likelihood method developed by \citet*{PPR00} to derive the 
He$^+$/H$^+$ ratio and some physical pro\-per\-ties of the ionized gas. In Table~\ref{helio} we include the 
He$^+$/H$^+$ ratios for each individual \ion{He}{1} line and the final adopted average value for each region 
$<$He$^+$/H$^+$$>$. In that table, we also include the corresponding value of $\tau_{3889}$ obtained from 
the maximum likelihood method as well as the corresponding $\chi^2$ parameter, which indicates a reasonable goodness 
of the fit obtained by the method in the four regions. 


\subsection{Ionic abundances from CELs
\label{ionicab}}

\begin{deluxetable}{l@{}c@{\hspace{10pt}}c@{\hspace{10pt}}c@{\hspace{10pt}}c@{\hspace{10pt}}
c@{\hspace{10pt}}c@{\hspace{10pt}}c@{\hspace{10pt}}} 
\rotate
\tabletypesize{\scriptsize}
\tablecaption{Ionic abundances from collisionally excited lines\label{ionic}}
\tablewidth{0pt}
\tablehead{
\colhead{12 + log(X$^{\rm m}$/H$^+$)} &
\multicolumn{2}{c}{(A) HII-2} &
\multicolumn{2}{c}{(B) HII-1} &
\multicolumn{2}{c}{(C) UV-1} &
\colhead{(D) UV-2} \\
 & \colhead{$t^2$=0.00} &
\colhead{$t^2$=0.072$\pm$0.027} & 
\colhead{$t^2$=0.00} &
\colhead{$t^2$=0.050$\pm$0.035} & 
\colhead{$t^2$=0.00} &
\colhead{$t^2$=0.061$\pm$0.024} &
\colhead{$t^2$=0.00}} 
\startdata
N$^{+}$	 	& 6.62 $\pm$ 0.04 	& 6.81 $\pm$ 0.10 	   	& 6.55 $\pm$ 0.04 	& 6.67 $\pm$ 0.12 	& 6.30 $\pm$ 0.05  	& 6.48 $\pm$ 0.10 	& 6.61 $\pm$ 
0.06   	\\
O$^{+}$	 	& 7.59 $\pm$ 0.06 	& 7.80 $\pm$ 0.11 	   	& 7.58 $\pm$ 0.05 	& 7.72 $\pm$ 0.12 	& 7.81 $\pm$ 0.06  	& 8.01 $\pm$ 0.11 	& 8.11 $\pm$ 
0.08   	\\
O$^{++}$ 	& 8.05 $\pm$ 0.03	& 8.34 $^{+0.15}_{-0.12}$	& 8.07 $\pm$ 0.03 	& 8.26 $^{+0.16}_{-0.13}$& 8.10 $\pm$ 0.03  	& 8.39 $\pm$ 0.14 	& 
7.87 $\pm$ 0.05   	\\
Ne$^{++}$	& 7.34 $\pm$ 0.05	& 7.65 $^{+0.16}_{-0.14}$	& 7.36 $\pm$ 0.05 	& 7.56 $^{+0.18}_{-0.15}$& 7.36 $\pm$ 0.05  	& 7.67 
$^{+0.17}_{-0.15}$& 7.15 $\pm$ 0.09  	\\
S$^{+}$ 	& 5.70 $\pm$ 0.05 	& 5.89 $\pm$ 0.10 	   	& 5.64 $\pm$ 0.05 	& 5.76 $^{+0.12}_{-0.10}$& 5.75 $\pm$ 0.05 	& 5.93 $\pm$ 0.09 	& 6.15 
$\pm$ 0.07   	\\
S$^{++}$ 	& 6.45 $\pm$ 0.05	& 6.77 $\pm$ 0.16 	   	& 6.44 $\pm$ 0.05 	& 6.64 $^{+0.18}_{-0.15}$& 6.45 $\pm$ 0.05 	& 6.76 $^{+0.17}_{-0.14}$& 
6.35 $\pm$ 0.07  	\\
Cl$^{+}$ 	& 4.07:	  		& 4.24: 			& 4.03: 		& 4.14: 		& \nodata	 	&  \nodata		& \nodata	    	\\
Cl$^{++}$	& 4.39 $\pm$ 0.08	& 4.67 $^{+0.16}_{-0.14}$	& 4.33 $\pm$ 0.06 	& 4.51 $^{+0.17}_{-0.14}$& 4.50 $\pm$ 0.07 	& 4.77 
$^{+0.16}_{-0.13}$& 4.19 $\pm$ 0.25  	\\
Cl$^{3+}$	& 3.73 $\pm$ 0.16	& 3.96 $\pm$ 0.19 		& 3.85 $\pm$ 0.09 	& 4.00 $\pm$ 0.15 	& 3.30 $\pm$ 0.17 	& 3.53 $\pm$ 0.20 	& \nodata 	    	
\\
Ar$^{++}$	& 5.93 $\pm$ 0.04	& 6.18 $^{+0.13}_{-0.11}$	& 5.90 $\pm$ 0.04 	& 6.06 $^{+0.14}_{-0.12}$& 5.92 $\pm$ 0.04 	& 6.16 $\pm$ 0.13 	& 
5.81 $\pm$ 0.07   	\\
Ar$^{3+}$	& 4.88 $\pm$ 0.06	& 5.18 $\pm$ 0.15 		& 5.00 $\pm$ 0.05 	& 5.19 $^{+0.17}_{-0.15}$& 4.34 $\pm$ 0.11 	& 4.63 $^{+0.18}_{-0.16}$& 
\nodata 	    	\\
Fe$^{+}$ 	& 4.53:	  		& 4.71: 			& 4.58: 		& 4.70: 		 & 4.62:	 	& 4.79:			& 4.96:	    	\\
Fe$^{++}$	& 5.53 $\pm$ 0.08	& 5.83 $\pm$ 0.16 	   	& 5.48 $\pm$ 0.08 	& 5.67 $^{+0.18}_{-0.16}$& 5.42 $\pm$ 0.11 	& 5.71 $^{+0.18}_{-0.16}$& 
5.89 $^{+0.11}_{-0.13}$ \\ 
\hline
log(N$^{+}$/O$^{+}$)    & --0.97 $\pm$ 0.07 	& --0.99 $\pm$0.14  		& --1.02 $\pm$ 0.07 	& --1.05 $\pm$0.16 	&  --1.51 $\pm$ 0.07 	& 
--1.53 
$\pm$0.14 	& --1.51 $\pm$ 0.10 	\\ 
\enddata
\end{deluxetable}

The IRAF package NEBULAR has been used to derive ionic abundances of O$^+$,  O$^{++}$, N$^+$, S$^+$, S$^{++}$, 
Ne$^{++}$, Ar$^{++}$, Ar$^{+3}$, Cl$^{++}$, and Cl$^{+3}$ for each region from the intensity of CELs. The electron 
densities and temperatures for the high and low ionization 
potential ions used are those corresponding to the aforementioned two-zone scheme. The finally adopted ionic 
abundances are listed in Table~\ref{ionic}, they correspond to the mean of the abundances derived from all 
the observed individual lines of each ion and weighted by their relative intensities. Errors were estimated from the 
uncertainties in electron density and temperature and those associated to the line intensity ratio with respect to
H$\beta$. Fe$^{++}$ abundances have been derived from 6 or 7 [\ion{Fe}{3}] lines except in the case of zone D, where 
only three lines were used. The lines selected for the derivation of the Fe$^{++}$/H$^+$ ratios are those not affected by line-blending and with line 
flux uncertainties less than 30\%. We have used a 34 level model atom that includes the collision strengths calculated by 
\citet{Z96} and the transition probabilities given by \citet{Q96}. Fe$^{++}$ abundances are also included in 
Table~\ref{ionic}. 

We have detected several [\ion{Fe}{2}] lines in our spectra, but they are severely affected by fluorescence effects
\citep{Ro99, verneretal00}. Unfortunately, the [\ion{Fe}{2}] $\lambda$8617 line, which is almost insensitive to fluorescence effects,
is not observed because it lies in one of our narrow observational  gaps. 
However, we have measured [\ion{Fe}{2}] $\lambda$7155,
a line which does not seem to be affected by fluorescence effects 
\citep{Ro96}. We have derived Fe$^+$ abundances from this line,
assuming that $I$(7155)/$I$(8617) $\sim$ 1 \citep{Ro96}, and using the 
calculations of \citet{bautistapradhan96}. The results obtained
imply low concentrations of Fe$^+$ (see Table~\ref{ionic}).
Due to the faintness of the 
[\ion{Fe}{2}] $\lambda$7155 line, and to the assumption
adopted, the derived Fe$^+$ abundances are only rough estimates and
will not be used in the Fe abundance determination.

We have detected [\ion{Cl}{2}] lines in two of the observed zones. However, the Cl$^{+}$/H$^{+}$
ratio cannot be derived from the NEBULAR routines because the atomic data of this ion is not included, instead we 
have used an old version of the five-level atom program of \citet{SD95} that is described by \citet*{DeRobertis87}. 
This version uses the atomic data for Cl$^{+}$ compiled by \citet{Mendoza83}, which are rather uncertain (Shaw 2003, 
personal communication). Therefore the Cl$^{+}$/H$^{+}$ ratio given in Table~\ref{ionic} should be interpreted as a 
rough approximation to the true one.

\subsection{Ionic abundances from RLs}

\begin{deluxetable}{lc@{\hspace{10pt}}c@{\hspace{10pt}}c@{\hspace{10pt}}c@{\hspace{10pt}}c@{\hspace{10pt}}c@{\hspace{10pt}}}
\tabletypesize{\scriptsize}
\tablecaption{O$^{++}$ and C$^{++}$ abundances from recombination lines$^{\rm a}$
\label{recom}}
\tablewidth{0pt}
\tablehead{ 
\colhead{ }&  
\multicolumn{2}{c}{(A) HII-2}  & 
\multicolumn{2}{c}{(B) HII-1} & 
\multicolumn{2}{c}{(C) UV 1}  \\ 
 & LTE  &  NLTE  &  LTE &  NLTE  & LTE & NLTE}
\startdata
O II $\lambda$4638.86 ........................... &
           38:  	&	20:	&    \nodata 	&  \nodata	&  47:		& 24:		\\
O II $\lambda$4641.81 ........................... &
           23:  	&	27:	&    \nodata 	& \nodata   	&  21 $\pm$ 7  	& 25 $\pm$ 8  	\\ 
O II $\lambda$4649.13 ........................... &
           12 $\pm$ 4 	&   25 $\pm$ 8  &   12 $\pm$ 4 	& 25 $\pm$ 8  	&  12 $\pm$ 4	& 29 $\pm$ 9	\\ 
O II $\lambda$4650.84 ........................... &
           34:	 	&       16:      &   36:	& 17:  		&  42:	 	& 18:		\\ 
O II $\lambda$4661.63 ........................... &
           21: 		&       12: 	&   20: 	& 12:          	&  47: 		& 26:      	\\
Sum value (all lines) ........... &
           20 		&       22 	&   17    	& 20          	&  24   	& 25      	\\
\noalign{\smallskip} 
{\bf O$\rm^{++}$/H$\rm^+$ adopted value} &
 		   \multicolumn{2}{c}{20 $\pm$ 8}   & \multicolumn{2}{c}{18 $\pm$ 7}  &  \multicolumn{2}{c}{24 $\pm$ 10} \\
\noalign{\smallskip}
\noalign{\smallskip}

 12+log(O$\rm^{++}$/H$\rm^+$) (RLs)& \multicolumn{2}{c}{8.30 $\pm$ 0.15} & \multicolumn{2}{c}{8.26 $\pm$ 0.15} &  
\multicolumn{2}{c}{8.38 $\pm$ 0.15}	  \\
 12+log(O$\rm^{++}$/H$\rm^+$) (CELs)  &\multicolumn{2}{c}{8.05 $\pm$ 0.03}  & \multicolumn{2}{c}{8.07 $\pm$ 0.03}    & 	 
\multicolumn{2}{c}{8.10 $\pm$ 0.03}   \\
\noalign{\smallskip}
\tableline
\noalign{\smallskip}
C II $\lambda$4267.15 ........................... &
\multicolumn{2}{c}{6 $\pm$ 2}   &\multicolumn{2}{c}{5:}      &\multicolumn{2}{c}{6  $\pm$ 3}    \\
\noalign{\smallskip} 
12+log(C$\rm^{++}$/H$\rm^+$) (RLs) & \multicolumn{2}{c}{7.82$^{+0.14}_{-0.22}$}    &     \multicolumn{2}{c}{7.75:}      
& \multicolumn{2}{c}{7.81$^{+0.15}_{-0.23}$}     \\  
12+log(C$\rm^{++}$/H$\rm^+$) (CELs)$^{\rm b}$ & \multicolumn{2}{c}{7.41$^{+0.16}_{-0.18}$}    &  \multicolumn{2}{c}{7.43 $\pm$ 
0.17}      & \multicolumn{2}{c}{7.48 $\pm$ 0.23}     \\  
 \noalign{\smallskip}
\noalign{\smallskip} 
 log(C$\rm^{++}$/O$\rm^{++}$) (RLs)&  \multicolumn{2}{c}{$-$0.48$^{+0.37}_{-0.41}$}  &   \multicolumn{2}{c}{$-0.51$:}        
& \multicolumn{2}{c}{$-$0.57$^{+0.33}_{-0.36}$}  \\  
 log(C$\rm^{++}$/O$\rm^{++}$) (CELs)$^{\rm b}$ &  \multicolumn{2}{c}{$-$0.62$^{+0.19}_{-0.20}$} &   
\multicolumn{2}{c}{$-$0.64$^{+0.16}_{-0.18}$} &  \multicolumn{2}{c}{$-$0.59$^{+0.25}_{-0.26}$}  \\ 
\enddata
\tablenotetext{a}{In units of 10$^{-5}$.}
\tablenotetext{b}{Value obtained from UV CELs (Kobulnicky et al. 1997).}
\end{deluxetable}

\begin{figure*}[t!]
\includegraphics[angle=270,width=1\linewidth]{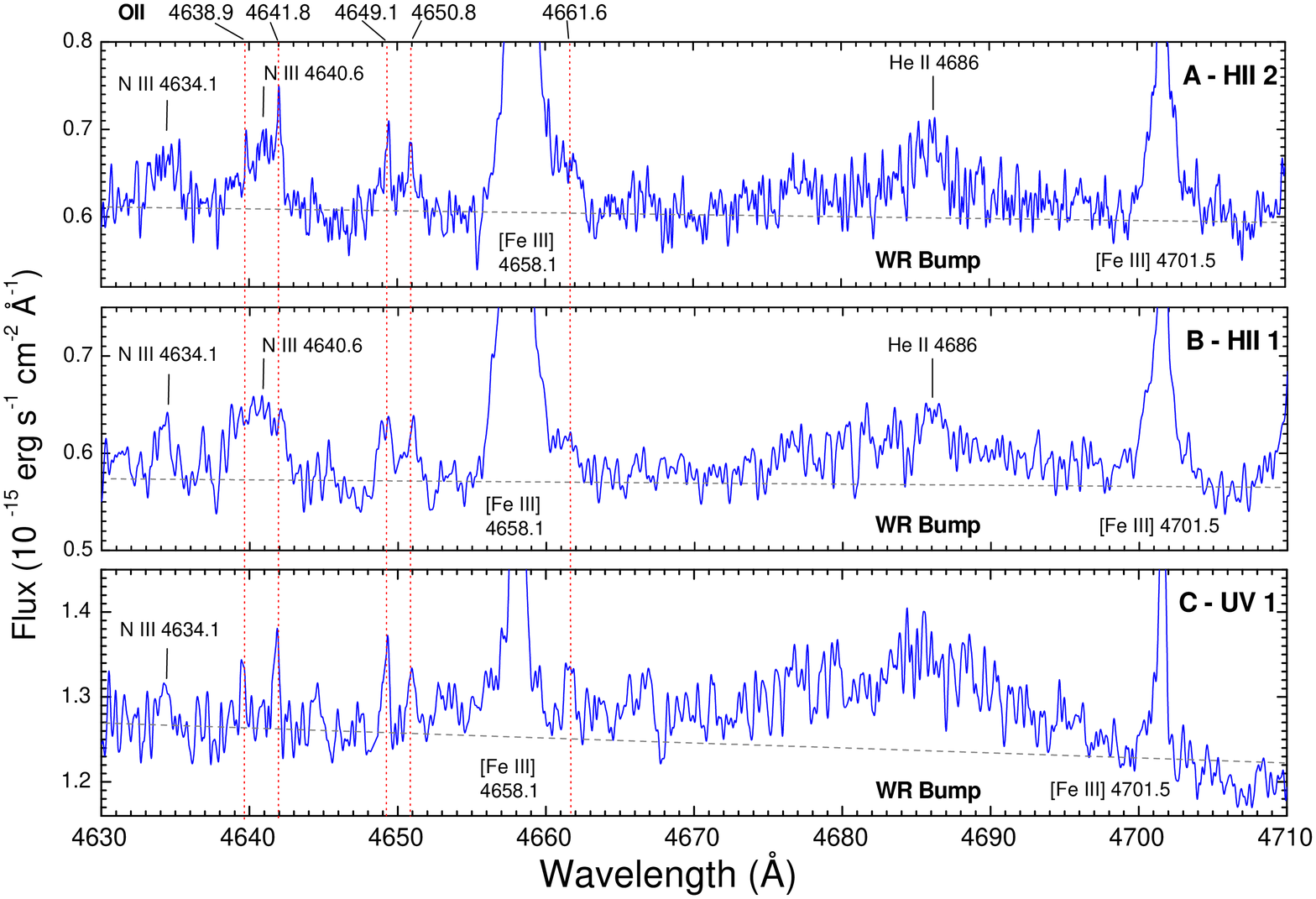}
\caption{\small{Sections of echelle spectra of zones A, B, and C of NGC 5253 showing the lines of 
multiplet 1 of \ion{O}{2} and the WR bump. Note the broad \ion{N}{3} $\lambda$4641 emission line blended with \ion{O}{2} 
$\lambda\lambda$4639,4642 emission lines in regions A and B. That line is absent in region C. A broad \ion{N}{3} $\lambda$4634 seems to be 
also present in region A.}}
\label{roii}
\end{figure*}

\begin{figure}[t!]
\includegraphics[angle=270,width=1\linewidth]{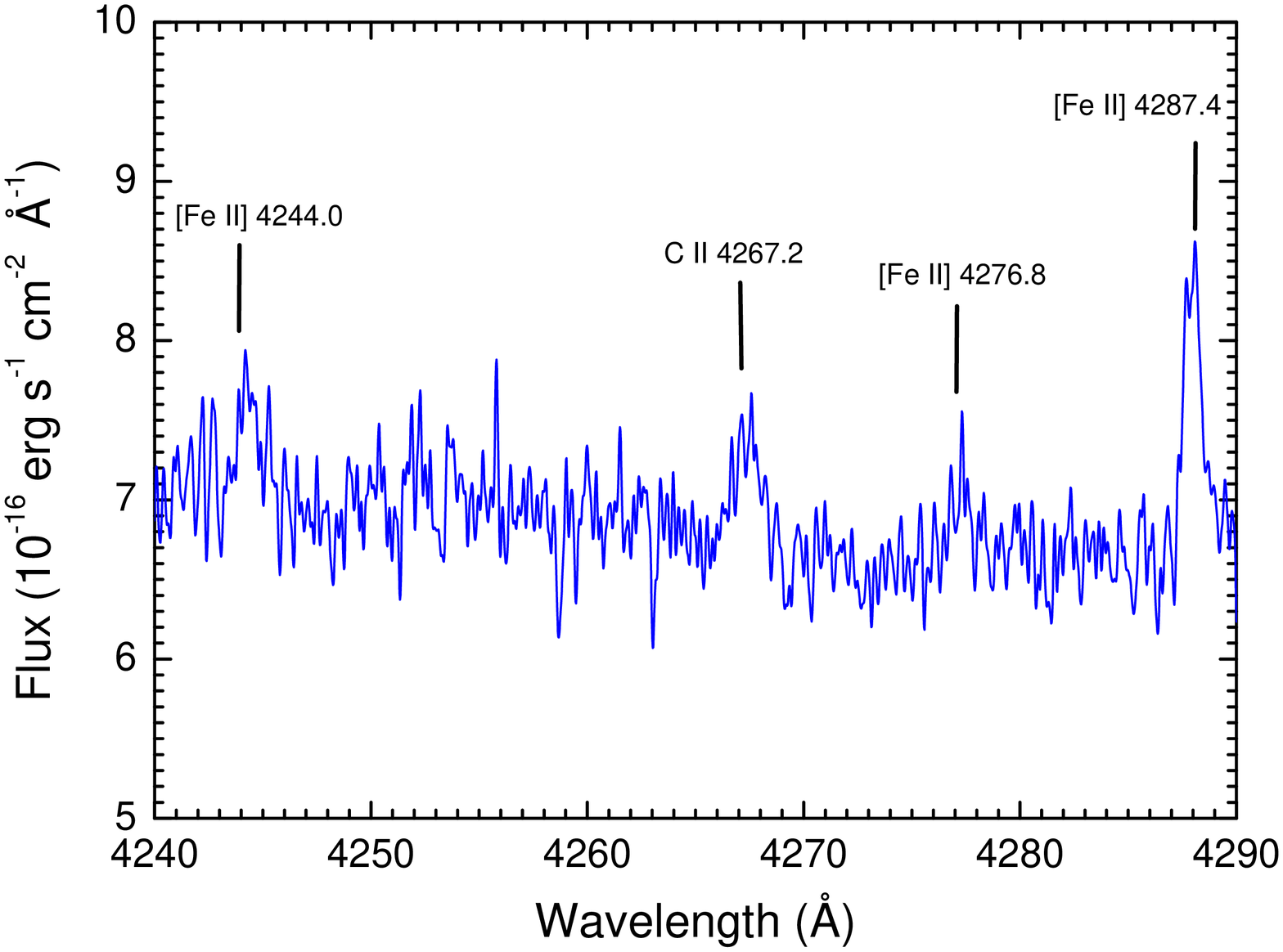}
\caption{\small{Section of echelle spectrum of region A of NGC 5253 showing the \ion{C}{2} $\lambda$4267 line.}}
\label{rcii}
\end{figure}

One of the main goals of this work has been the measurement of RLs of heavy element ions in NGC 5253, the first dwarf 
starburst galaxy where these kinds of lines have been unambiguously detected. Three of the four zones observed show the presence of  
RLs of heavy element ions in their spectra. These lines are those belonging to multiplet 1 of \ion{O}{2} (see 
Fi\-gu\-re~\ref{roii}) and \ion{C}{2} $\lambda$4267 (see Figure~\ref{rcii}). All these lines are produced by pure 
recombination [see \citet{E98,E04}, and references therein] and their intensities depend weakly on electron 
temperature and density. We have computed the abundances assuming the adopted values of $n_e$ and $T_e$(High) given 
in Table~\ref{temden} for each zone. The atomic data used and the methodology for the derivation of the abundances 
from RLs are the same as in \citet{GRE04}. The lines of multiplet 1 of \ion{O}{2} are not in LTE for densities 
$n_e<$10000  cm$^{-3}$ \citep{Ruiz03}. We have used the prescriptions given by \citet*{PPR05} to calculate the 
appropriate corrections for the abundances obtained from individual \ion{O}{2} lines. These corrected abundances show 
very good agreement with those obtained using the sum of the intensities of all the lines of the multiplet, which is 
not affected by non LTE effects. Table~\ref{recom} shows the O$^{++}$/H$^+$ and C$^{++}$/H$^+$ ratios obtained from
individual RLs as well as the correction for non LTE effects and the corresponding sum values for the \ion{O}{2} 
lines. In that table we also compare with the same ratios obtained from CELs. 
In the case of the C$^{++}$/H$^+$ ratio, we have compared our determinations based on RLs with those obtained by 
\citet{Kobulnicky97} from UV CELs for exactly the same zones but with a slightly smaller aperture. Consistently with the result found for 
O$^{++}$/H$^+$, the C$^{++}$/H$^+$ 
ratios derived from RLs are systematically larger than those derived from CELs. 

\section{Abundance Discrepancy \\ and Temperature Fluctuations}

\begin{deluxetable}{c@{\hspace{10pt}}c@{\hspace{10pt}}c@{\hspace{10pt}}c@{\hspace{10pt}}}
\tabletypesize{\scriptsize}
\tablecaption{NGC~5253 $t^2$ parameter
\label{t2}}
\tablewidth{0pt}
\tablehead{
\colhead{Method} &
\multicolumn{3}{c}{$t^2$} \\
& \colhead{(A) HII-2} & 
\colhead{(B) HII-1} &
\colhead{(C) UV-1} }
\startdata
O$^{\rm ++}$ (R/C)	& 0.064 $\pm$ 0.035	       	& 0.050 $\pm$ 0.035     & 0.060 $\pm$ 0.030	      	       \\
C$^{\rm ++}$ (R/C)	& 0.084 $^{+0.027}_{-0.049}$ 	& 0.073:	       	& 0.062 $^{+0.027}_{-0.052}$           \\
Adopted 		& 0.072 $\pm$ 0.027		& 0.050 $\pm$ 0.035 	& 0.061 $\pm$ 0.024 		  \\ 
\enddata
\end{deluxetable}

As it can be seen in Table~\ref{recom}, the discrepancy between ionic abundances obtained from RLs and CELs are between 0.19 and 
0.28 dex in the case of O$^{++}$/H$^+$ and between 0.30 and 0.40 dex in the case of C$^{++}$/H$^+$. Very similar values 
have been obtained for different Galactic and extragalactic \ion{H}{2} regions. \citet*{TP80} proposed that the 
abundance discrepancy could be related to the presence of spatial temperature fluctuations because of the different 
functional temperature dependence of RLs and CELs. Table~\ref{t2} shows the different determinations of $t^2$ and 
our final adopted value. The two upper rows of the table show the $t^2$ parameters that make the O$^{++}$ and 
C$^{++}$ abundances obtained from RLs and CELs to coincide [denoted as O$^{++}$(C/R) and C$^{++}$(C/R) in the table]. 
We can see that these two $t^2$ values are consistent within the errors for all the zones. The final adopted 
value of $t^2$ is the weighted average of the individual determinations for each zone. 

\section{Total Abundances}

Total abundances have been determined for He, C, N, O, Ne, S, Cl, Ar, and Fe and are shown in Table~\ref{totabun}. 
The non-detection of the nebular \ion{He}{2} $\lambda$4686 line in our deep spectra implies the absence of a 
significant amount of He$^{++}$ and O$^{3+}$ in the ionized gas of NGC 5253. Then, the 
use of the relation O/H = O$^+$/H$^+$ + O$^{++}$/H$^+$ is entirely justified. For the rest of the elements, we have 
to adopt a set of ioni\-zation correction factors (ICFs) to correct for the unseen ionization stages. 

For carbon we have adopted the ICF derived from photoionization models by \citet{G99}. This correction seems to be 
fairly appropriate considering the high ionization degree of the three brightest zones of NGC 5253 where the \ion{C}{2} 
line has been detected. 

In the case of neon we have applied the classical ICF proposed by \citet{PC69}, that assumes that the ionization 
structure of Ne is similar to that of O. This is a very good approximation for high ionization degree objects, where 
a small fraction of Ne$^+$ is expected. \citet*{Martin-Hernandez05}, using $N$-band NIR spectroscopy, have 
estimated 12+log(Ne$^+$/H$^+$)=6.46, for a zone located between our regions A and B. Assuming this Ne$^+$ abundance 
for A and B we obtain an ICF of $\sim$1.12, somewhat lower than the value of $\sim$1.36 that gives our adopted ICF 
scheme.  

We have measured two ionization stages of S, S$^+$ and S$^{++}$, in all regions, but a significant contribution of 
S$^{3+}$ is expected. We have adopted the ICF given by \citet{S78}, which is based on photoionization models of 
\ion{H}{2} regions and is expressed as a function of the O$^+$/O ratio.

We have measured lines of all possible ionization stages of chlorine in two of the zones (A and B). In any case, as 
it can be seen in Table~\ref{ionic}, the dominant ionization stage is Cl$^{++}$ and the contribution of Cl$^+$ to the 
total abundance is rather small, at least in the high ionization degree regions as A, B, and C. Hence, we 
have neglected any contribution of the singly ionized stage for the determination of the total abundance of chlorine 
in zone C. In the case of region D, its lower ionization recommends to take into account the Cl$^+$ fraction, for 
that, we have adopted the relation by \citet{PTP77}. 

For argon we have determinations of the Ar$^{++}$ and Ar$^{3+}$ abundances. We find
that Ar$^{++}$ is almost one order of magnitude more abundant than Ar$^{3+}$, indicating that most Ar is in the form 
of Ar$^{++}$. However, some contribution of Ar$^{+}$ is expected.  \citet*{Martin-Hernandez05} have obtained an upper 
limit for Ar$^{+}$/Ar$^{++}$ $<$ 0.35 for a zone located between our regions A and B. In our case, we have adopted 
the ICF of \citet*{ITL94}.

For Fe, we have measured lines of two stages of ionization of iron: Fe$^{+}$ and Fe$^{++}$, but we expect an 
important contribution of Fe$^{3+}$. Since the Fe$^+$ contribution is small and uncertain, (see \S~\ref{ionicab}) 
the total iron abundance have been obtained from the Fe$^{++}$/H$^+$ ratio and the ICF obtained by \citet{RR05}.

\begin{deluxetable}{l@{}c@{\hspace{10pt}}c@{\hspace{10pt}}c@{\hspace{10pt}}c@{\hspace{10pt}}
c@{\hspace{10pt}}c@{\hspace{10pt}}c@{\hspace{10pt}}} 
\rotate
\tabletypesize{\tiny}
\tablecaption{Total abundances and abundance ratios
\label{totabun}}
\tablewidth{0pt}
\tablehead{
\colhead{12 + log(X/H)} &
\multicolumn{2}{c}{(A) HII-2} &
\multicolumn{2}{c}{(B) HII-1} &
\multicolumn{2}{c}{(C) UV-1} &
\colhead{(D) UV-2} \\
 & \colhead{$t^2$=0.00} &
\colhead{$t^2$=0.072$\pm$0.027} & 
\colhead{$t^2$=0.00} &
\colhead{$t^2$=0.050$\pm$0.035} & 
\colhead{$t^2$=0.00} &
\colhead{$t^2$=0.061$\pm$0.024} &
\colhead{$t^2$=0.00}} 
\startdata
He$^{\rm a}$	& 10.97 $\pm$ 0.02	& 10.95 $\pm$ 0.02 	& 10.95 $\pm$ 0.01 	& 10.94 $\pm$ 0.02 	& 10.93 $\pm$ 0.02  	& 10.91 $\pm$ 0.02 	& 11.08 
$\pm$ 0.06  	\\
He$^{\rm b}$	& 10.92 		& 10.92 		& 10.91	 		& 10.91 		& 10.87 		& 10.87 		& 10.88  		\\
C		& 7.92 $\pm$ 0.15 	& 7.92 $\pm$ 0.15 	& 7.84: 		& 7.84: 		& 7.92 $\pm$ 0.16 	& 7.92 $\pm$ 0.16 	& \nodata 		\\
N$^{\rm c}$	& 7.21 $\pm$ 0.07 	& 7.47 $\pm$ 0.17 	& 7.16 $\pm$ 0.07 	& 7.33 $\pm$ 0.18 	& 6.77 $\pm$ 0.08  	& 7.01 $\pm$ 0.17 	& 6.81 $\pm$ 0.11  	
\\
N$^{\rm d}$	& 7.27 $\pm$ 0.06 	& 7.54 $\pm$ 0.14 	& 7.23 $\pm$ 0.06 	& 7.40 $\pm$ 0.15 	& 6.83 $\pm$ 0.07 	& 7.08 $\pm$ 0.15 	& 6.82 $\pm$ 0.07  	
\\
N$^{\rm e}$	& 7.32 			& 7.61 			& 7.25 			& 7.50 			& 6.93 			& 7.25 			& 6.88 			\\
O$^{\rm f}$ 	& 8.18 $\pm$ 0.04   	& 8.45 $\pm$ 0.12 	& 8.19 $\pm$ 0.04 	& 8.37 $\pm$ 0.13 	& 8.28 $\pm$ 0.04  	& 8.54 $\pm$ 0.11 	& 8.31 
$\pm$ 0.07  	\\
O$^{\rm g}$	& 8.42 $\pm$ 0.13  	& 8.42 $\pm$ 0.13 	& 8.37 $\pm$ 0.10 	& 8.37 $\pm$ 0.10 	& 8.53 $\pm$ 0.09  	& 8.53 $\pm$ 0.09 	& \nodata 	   	
\\
Ne 		& 7.47 $\pm$ 0.06   	& 7.76 $\pm$ 0.20 	& 7.49 $\pm$ 0.06 	& 7.67 $\pm$ 0.22 	& 7.54 $\pm$ 0.07  	& 7.82 $\pm$ 0.20 	& 7.61 $\pm$ 0.09  	
\\
S 		& 6.60 $\pm$ 0.07 	& 6.91 $\pm$ 0.15 	& 6.59 $\pm$ 0.05 	& 6.79 $\pm$ 0.13 	& 6.58 $\pm$ 0.05  	& 6.88 $\pm$ 0.12 	& 6.57 $\pm$ 0.07  	
\\
Cl$^{\rm h}$	& 4.62 $\pm$ 0.08	& 4.86 $\pm$ 0.13 	& 4.59 $\pm$ 0.07 	& 4.75 $\pm$ 0.13 	& \nodata 		& \nodata 		& \nodata		\\
Cl$^{\rm i}$	& 4.53 $\pm$ 0.07	& 4.79 $\pm$ 0.10 	& 4.49 $\pm$ 0.06 	& 4.66 $\pm$ 0.12 	& 4.59 $\pm$ 0.08  	& 4.85 $\pm$ 0.14 	& \nodata 		
\\
Ar 		& 5.99 $\pm$ 0.04 	& 6.24 $\pm$ 0.10 	& 5.97 $\pm$ 0.04 	& 6.14 $\pm$ 0.12 	& 5.98 $\pm$ 0.04  	& 6.21 $\pm$ 0.12 	& 6.01 $\pm$ 0.07  	
\\
Fe 		& 6.08 $\pm$ 0.11 	& 6.43 $\pm$ 0.23 	& 6.01 $\pm$ 0.11 	& 6.23 $\pm$ 0.27 	& 5.82 $\pm$ 0.14  	& 6.17 $\pm$ 0.25 	& 6.06 $\pm$ 0.15  	
\\ \hline
log(X/O) 	& 			& 			& 			& 			& 			& 			& 			\\ \hline
C/O		& \nodata 		& --0.53 $\pm$ 0.18 	& \nodata 		& --0.53: 		& \nodata 		& --0.62 $\pm$ 0.19 	& \nodata 		\\
N/O$^{\rm j}$	& --0.91 $\pm$ 0.07 	& --0.91 $\pm$ 0.18  	& --1.02 $\pm$ 0.07 	& --0.97 $\pm$ 0.19 	& --1.50 $\pm$ 0.08 	& --1.46 
$\pm$ 0.18 	& --1.49 $\pm$ 0.10 	\\
S/O    		& --1.58 $\pm$ 0.08 	& --1.54 $\pm$ 0.18 	& --1.60 $\pm$ 0.08 	& --1.58 $\pm$ 0.17 	& --1.69 $\pm$ 0.09 	& --1.66 $\pm$ 
0.16 	& --1.74 $\pm$ 0.13 	\\
Ne/O   		& --0.71 $\pm$ 0.08 	& --0.69 $\pm$ 0.22 	& --0.70 $\pm$ 0.08 	& --0.70 $\pm$ 0.24 	& --0.74 $\pm$ 0.08 	& --0.72 $\pm$ 
0.22 	& --0.70 $\pm$ 0.15 	\\
Cl/O$^{\rm k}$	& --3.65 $\pm$ 0.08 	& --3.66 $\pm$ 0.15 	& --3.70 $\pm$ 0.08 	& --3.71 $\pm$ 0.17 	& --3.68 $\pm$ 0.09 	& --3.69 
$\pm$ 0.17 	&  \nodata 		\\
Ar/O   		& --2.19 $\pm$ 0.07 	& --2.21 $\pm$ 0.15 	& --2.21 $\pm$ 0.07 	& --2.23 $\pm$ 0.17 	& --2.30 $\pm$ 0.08 	& --2.33 $\pm$ 
0.16 	& --2.30 $\pm$ 0.13 	\\
Fe/O   		& --2.10 $\pm$ 0.12 	& --2.02 $\pm$ 0.25 	& --2.18 $\pm$ 0.11 	& --2.14 $\pm$ 0.29 	& --2.46 $\pm$ 0.14 	& --2.37 $\pm$ 
0.26 	& --2.25 $\pm$ 0.16 	\\ 
\enddata
\tablenotetext{a}{ICF from \citet*{PTPR92}.}
\tablenotetext{b}{ICF from \citet{S90}.}
\tablenotetext{c}{ICF from \citet{PC69}.}
\tablenotetext{d}{ICF from \citet{Mathis91}.}
\tablenotetext{e}{ICF from \citet*{Moore04}.}
\tablenotetext{f}{Based on ionic abundances derived from CELs.}
\tablenotetext{g}{O$^{++}$/H$^+$ from RLs and O$^+$/H$^+$ from CELs and t$^2$.}
\tablenotetext{h}{Cl/H = Cl$^{+}$/H$^+$ + Cl$^{++}$/H$^+$ + Cl$^{+3}$/H$^+$.}
\tablenotetext{i}{ICF from \citet{PTP77}.}
\tablenotetext{j}{Assuming the ICF(N) from \citet{Mathis91}.}
\tablenotetext{k}{Assuming the ICF(Cl) from \citet{PTP77}.}
\end{deluxetable}

As it was said in the introduction, some previous works have found a remarkable nitrogen enrichment --and perhaps 
also helium \citep*{CTM86}-- in some zones of NGC 5253. Therefore a special care has been taken for the determination 
of the total abundance of these two elements. For helium, the aforementioned absence of a measurable intensity of the
\ion{He}{2} $\lambda$4686 line indicates that He$^{++}$ is not important in all the zones observed. However, we have 
to include a correction for the presence of neutral helium inside the nebulae. In any case, the high ionization 
degree of regions A, B, and C implies that the correction for neutral helium should be small. We have used the 
empirical ICF proposed by \citet*{PTPR92} for all the different zones, although it is expected to be a good 
approximation only for those zones with high ionization degree. In the case of zone D, we find that the total helium abundance is 
higher than in the other regions. This may be a spurious result due to several reasons: a) this zone has 
comparatively much larger line intensity uncertainties and fewer \ion{He}{1} lines available, and b) the ionization 
degree of this zone is the lowest of the observed regions, and the ICF for helium is comparatively more uncertain. 
Therefore, we will no longer 
consider the He/H ratio of region D in our discussion. For comparison, we have also used the 
photoionization models by \citet{S90} to estimate the ICF for helium in regions A, B, and C, finding that the 
appropriate models for the properties of those zones (models C2C1 and C2D1) give very small values of the ICF ($\sim$1.02). 

In Table~\ref{totabun} we include estimations of the N abundance based on three different ICF schemes. The first row 
includes the values assuming the standard ICF factor by \citet{PC69}: N/O = N$^+$/O$^+$, which is a reasonably good 
approximation for the excitation degree of the observed regions of NGC 5253. The second row includes the total 
abundances obtained from the formulae provided by \citet{Mathis91}, which give abundances between 0.02 and 0.07 dex 
higher than the ICF of Peimbert \& Costero. The third estimation has been obtained using the results of the 
photoionization models by \citet*{Moore04}. This last determination gives the highest values of N/H, about 0.08 and 
0.16 dex higher than the ICF of Peimbert \& Costero. 

In Table~\ref{totabun} we show the total abundances for the different zones of NGC 5253 observed for $t^2$=0.00 and 
the $t^2$ value adopted for each object given in Table~\ref{t2}. Our abundances derived from CELs and $t^2$=0 are in 
good agreement with previous determinations by \citet{WR89} and \citet{Kobulnicky97}. We confirm the remarkable 
difference in the nitrogen abundance between different areas of the nebula: regions A and B show higher N/H ratios 
than regions C and D. This difference seems to be real because it cannot be accounted for the observational 
uncertainties, it does not depend on the ICF scheme used for N, and it is independent on the assumption or not of the 
temperature fluctuations paradigm. On the other hand, the O/H ratio seems to be slightly higher in zones C and D with 
respect to the other two zones, $\sim$0.10 dex in the case of CELs and $\sim$0.13 dex in the case of RLs. This difference has not been 
reported in previous works and is of the order of the estimated uncertainties. There is also a slight difference of the He/H 
ratio in zones A, B, and C. If this difference is real, it would confirm the localized helium enrichment found by 
\citet{CTM86} in a region encompassing our zones A and B, but with much lower values of the abundance difference in our case. 

\section{Further analysis of the \\ emission line profiles}

\begin{deluxetable}{l@{} c@{\hspace{10pt}} c@{\hspace{10pt}} c@{\hspace{10pt}} c@{\hspace{10pt}} c@{\hspace{10pt}} c@{\hspace{10pt}} 
c@{\hspace{10pt}}  c@{\hspace{10pt}} }
\rotate
\tabletypesize{\scriptsize}
\tablecaption{Line intensities$^{\rm a}$, physical conditions, and abundances$^{\rm b}$ of the velocity components \label{componentes}}
\tablewidth{0pt}
\tablehead{
\colhead{ } &
\multicolumn{2}{c}{(A) HII-2} &
\multicolumn{2}{c}{(B) HII-1} &
\multicolumn{2}{c}{(C) UV-1} &
\multicolumn{2}{c}{(D) UV-2} \\
                 &
\colhead{Narrow} &
\colhead{Broad}  & 
\colhead{Narrow} &
\colhead{Broad}  & 
\colhead{Narrow} &
\colhead{Broad}  &
\colhead{Narrow} &
\colhead{Broad}  } 
\startdata

[O II] 3726 &  41 &  69 &  74 &  54 &  78 & 101 & 112 & 144 \\

[O II] 3729 &  47 &  75 &  97 &  54 &  99 & 130 & 166 & 204 \\

[O III] 4363 &   4 &   7 &   5 &   7 &   4 &   4 &   3 & 3 \\

 \Hb\  4861 & 100 & 100 & 100 & 100 & 100 & 100 & 100 & 100 \\
 
[O III] 5007 & 507 & 638 & 506 & 623 & 488 & 458 & 333 & 305 \\

   \Ha\  6563 & 282 & 282 & 282 & 282 & 282 & 282 & 282 & 282   \\
   
 [N II] 6583 & 24 & 33 & 16 &  30 &  11 & 16 & 18 & 25   \\
 
 [S II] 6716 & 12 & 15 & 13 & 11 & 11 & 25 & 38 & 33   \\
 
 [S II] 6731 & 11 & 14 & 11 & 11 & 10 & 21 & 27 & 29 \\
	\noalign{\smallskip}
\hline	                
	\noalign{\smallskip}   
$C$(\Hb) & 0.29 & 0.18   &   0.42  &   0.30   & 0.37 & 0.12 & 0.46 & 0.05   \\
$n_e$ (cm$-3$) & 510 & 610 & 380 & 760 & 380& 350 & 170 & 320 \\
$T_e$(O III) (K) & 11800 & 12100 & 11600 & 12300 & 10800 & 11100 & 11200 & 10800   \\
$T_e$(O II)$^{\rm c}$ (K) &  11300 & 11400 & 11100 & 11600 & 10500 & 10800 & 10800 & 10600 \\
12 +log(O$^{++}$/H$^+$) & 8.14 & 8.09 & 8.04 & 8.06 & 8.13 & 8.05 & 7.91 & 7.92 \\
log(O$^{++}$/O$^+$) & 0.64 & 0.52 & 0.36 & 0.62 & 0.34 & 0.19 & $-$0.01 & $-$0.15 \\
log (N$^+$/O$^+$) &  $-$0.87 & $-$0.90 & $-$1.31 & $-$0.83 & $-$1.53 & $-$1.44 & $-$1.47 & $-$1.45   \\
12+log(O/H) & 8.23 & 8.21 & 8.20 & 8.15 & 8.30 & 8.27 & 8.22 & 8.31 \\
\enddata
\tablenotetext{a}{Dereddened line intensity ratios with respect to I(H$\beta$)=100.}
\tablenotetext{b}{Determined from the intensity of collisionally excited lines.} 
\tablenotetext{c}{Determined from $T_e$(O III) using the \citet{G92} calibration.} 
\end{deluxetable}

In \S 4 we have shown the presence of different velocity components in the emission line profiles of the four zones studied in NGC 5253. We have 
carried out a further analysis of those components, estimating their physical conditions and chemical abundances. We have performed a double 
Gaussian fit to the profiles of the main emission lines making use of the Starlink DIPSO software \citep{HM90}. The selected lines were:  
[\ion{O}{2}] $\lambda\lambda$3726, 3729, [\ion{O}{3}] $\lambda$$\lambda$4363, 5007, \Hb, \Ha, [\ion{N}{2}] 
$\lambda$$\lambda$5755, 6583, and [\ion{S}{2}] $\lambda\lambda$6716, 6731. In 
Table~\ref{componentes} we indicate the emission line intensities for the two components in each region as well as the reddening coefficient (derived 
from the \Ha/\Hb\ ratio), the phy\-sical conditions [$n_e$, $T_e$(\ion{O}{3}) and $T_e$(\ion{O}{2}), that was determined using the \citet{G92} 
calibration between $T_e$(\ion{O}{3}) and $T_e$(\ion{O}{2})], some ionic abundances derived from CELs (O$^+$, 
O$^{++}$ and N$^+$), and the O/H and N$^+$/O$^+$ ratios. From the table, we can see that most of the values of the different parameters are very 
similar for 
the narrow and broad component of each zone, and also similar to the values obtained for the integrated line profile. However, we have found some 
interesting exceptions. First, the reddening coefficient seems to be somewhat lower in the case of the broad components in all cases. Secondly, 
there is a difference in the N/O ratio of the broad and narrow components of zone B. The narrow component has a rather normal ratio for a galaxy of 
its corresponding O abundance, and the broad one shows a higher nitrogen abundance. This would indicate that the broad component contains the 
localized N enrichment. However, this behaviour is not observed in zone A. Here both components show a high N/O, similar to that reported for the 
integrated profile.

\section{Absorption lines and \\ stellar kinematics}

\begin{figure}[t!]
\includegraphics[angle=270,width=1\linewidth]{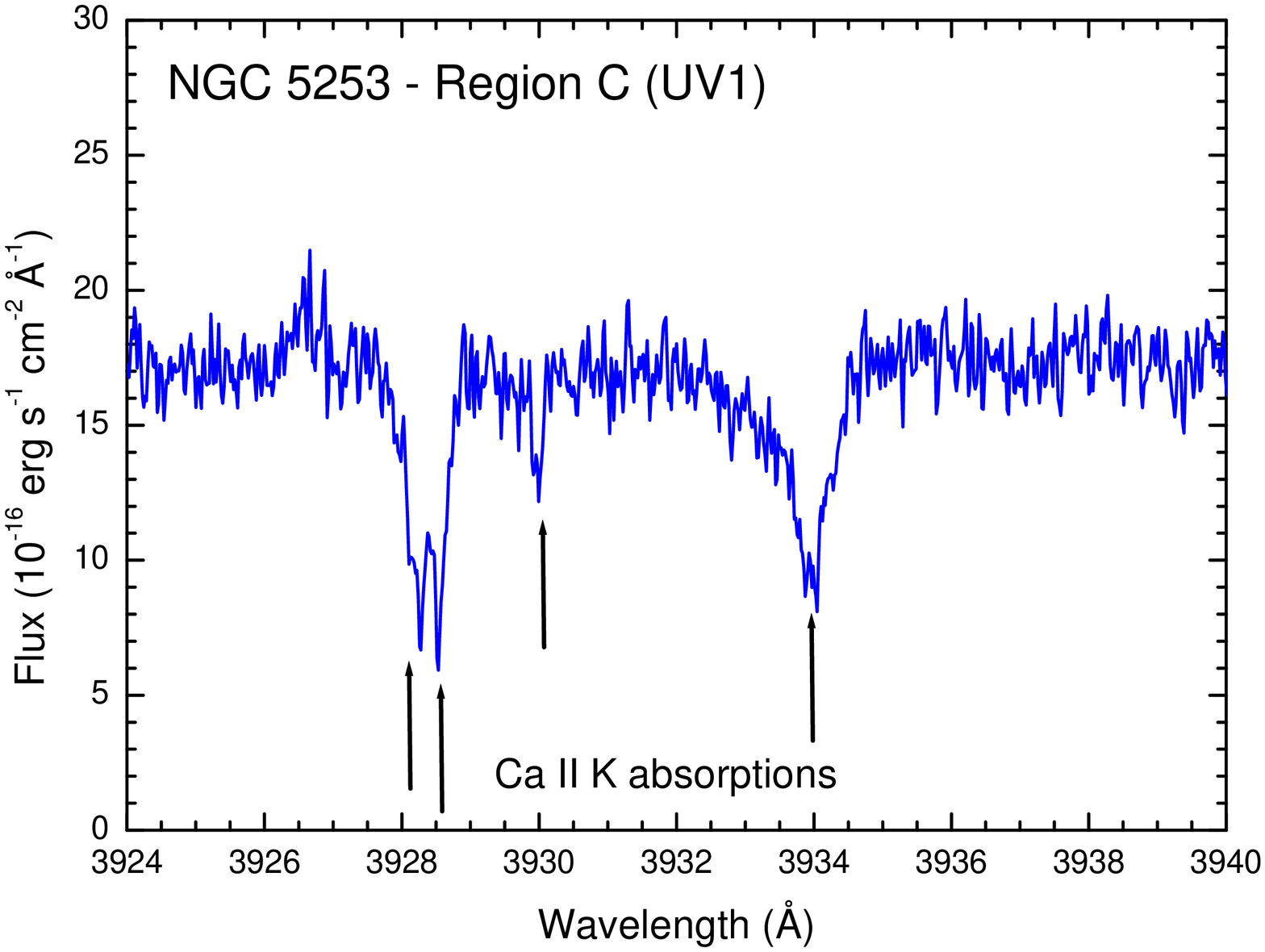}
\caption{\small{\ion{Ca}{2} K $\lambda$3933 absorptions for region C of NGC 5253. Note the 
existence of several components. The wavelength axis has been corrected for the recession velocity of 
the galaxy. The two Galactic components are the bluest ones.}}
\label{absor1}
\end{figure}

Although our main analysis of the spectra of NGC 5253 is focused on emission lines, we report the presence of some absorption features. 
They have been identified as \ion{Ca}{2} K $\lambda$3933 and H $\lambda$3968, \ion{Mg}{1} $\lambda$5167, 
\ion{Na}{1} $\lambda\lambda$5890, 5896 and \ion{Ca}{2} $\lambda\lambda$8498, 8662. The strongest absorption lines are 
\ion{Ca}{2} H, K, which are unambiguously detected in all regions (see Figure~\ref{absor1} for the example of region C). We find that the
\ion{Ca}{2} H, K lines are splitted into several kinematical components (at least four), but two of them are due to Galactic absorption systems 
because of their low heliocentric radial velocity ($v_{rad}\sim$0 and $\sim$20 \mbox{km s$^{-1}$}, respectively).

The broader absorption component shows a radial velocity rather similar to that of the ionized gas 
of each region, and can be interpreted as the contribution of the stellar populations of NGC 5253. A second and narrower 
component with radial 
velocity $\sim$+95 km s$^{-1}$ is also detected in all regions. It might be of Galactic origin because NGC 5253 is 
located in the fourth quadrant of the Galaxy and radial velocities of the order of +100 km s$^{-1}$ are common. 
However, the relatively large galactic latitude, $b\sim +30^{\circ}$, makes rather unlikely its true Galactic nature. 
Other possibilities are that this absorption feature is produced by a relatively nearby high-velocity 
cloud or by an intergalactic cloud located at $\sim$1.3 Mpc. Using the formulae provided by \citet{Smoker05}, we find that the 
column density of this component is around 17 times smaller than that associated with NGC 5253.

\begin{figure}[t!]
\includegraphics[angle=270,width=1\linewidth]{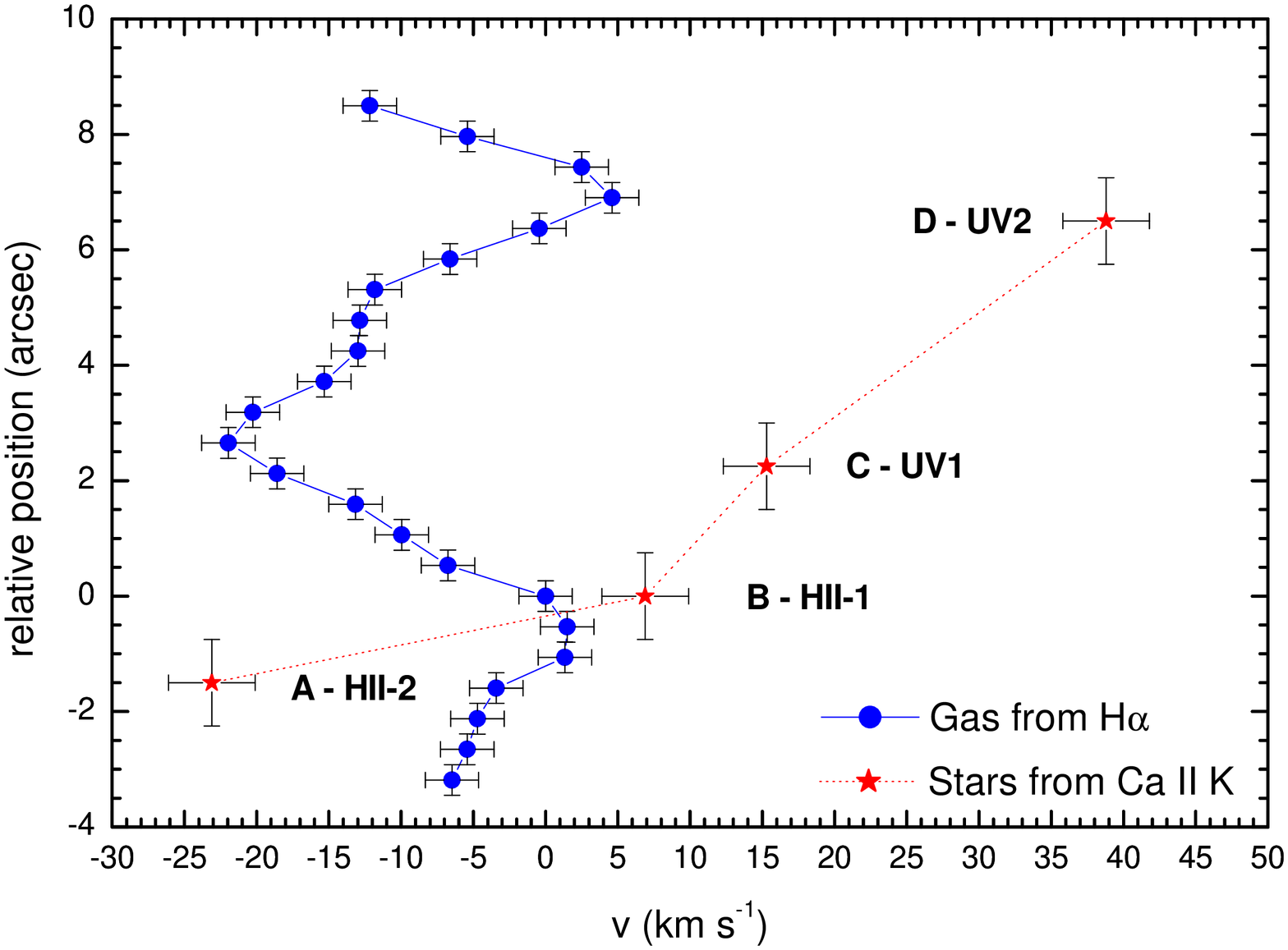}
\caption{\small{Position-velocity diagram of the \Ha\ emission line compared to the stellar component kinematics 
derived from the \ion{Ca}{2} K absorption features observed in each region.}}
\label{absor2}
\end{figure}

Data of radial velocities derived from the \ion{Ca}{2} K absorption feature associated with NGC 5253 have been included in Figure~\ref{absor2} 
to compare the kinematics of the ionized gas with that of the stellar component. Notice that 
$v_{rad}$ determined from \ion{Mg}{1}, \ion{Na}{1}, and red \ion{Ca}{2} absorptions detected in C and D are in good agreement with those
derived from \ion{Ca}{2} K. Although we only considered 4 points for constructing the position-velocity dia\-gram of the absorption features, 
it is clear that gas and stars are kinematically 
decoupled. The stellar component seems to follow a behaviour consistent with a rotation pattern. If these velocity differences between gas and stars 
are due to the presence of a supershell, its velocity relative to the stellar velocity 
is $\sim$40--50 km s$^{-1}$. This value agrees with the expansion velocity of the supershell detected by \citet{Marlowe95} in NGC 5253.

\section{The age of the bursts and \\ the massive stellar population}

\begin{deluxetable}{lcccc}
\tabletypesize{\footnotesize}
\tablecaption{Analysis of O and WR populations in observed knots of NGC 5253.\label{WR}} 
\tablewidth{0pt}
\tablehead{

& \colhead{(A) HII-2} & 
\colhead{(B) HII-1} &
\colhead{(C) UV-1} &
\colhead{(D) UV-2}}
\startdata

    $F$ (WR Bump) (10$^{-16}$ erg s$^{-1}$ cm$^{-2}$)  & 11 $\pm$ 3  &  10 $\pm$ 2   & 65 $\pm$ 18               
&  1.2 $\pm$ 0.7\\       
	$L$ (WR Bump) (10$^{36}$ erg s$^{-1}$) &  2.25 $\pm$ 0.66  &  2.01 $\pm$ 0.46  &  13.1 $\pm$ 3.6    &   
0.24 $\pm$ 0.14 \\
	WNL eq. stars$^{\rm a}$     &   1.3 $\pm$ 0.4  & 1.2 $\pm$ 0.3 & 8 $\pm$ 2 & 0.14 $\pm$ 0.08 \\
	$L$ (H$\beta$) (10$^{36}$ erg s$^{-1}$)  & 271 $\pm$ 9  & 272 $\pm$ 8 &  204 $\pm$ 7  & 51.6 $\pm$ 1.9  \\  
	O7V stars$^{\rm b}$  &   57 $\pm$ 2  &  57 $\pm$ 2 &  43 $\pm$ 1 & 10.8 $\pm$ 0.4 \\
	Adopted Age (Myr)  &  3.0 $\pm$ 0.1  & 2.7 $\pm$ 0.1 & 4.6 $\pm$ 0.1 & 5.1 $\pm$ 0.1\\
	$\eta$$^{\rm c}$     &  $\sim$1.0 & $\sim$1.1 & $\sim$0.4 & $\sim$0.3 \\
	O total stars      &  59 $\pm$ 2 &  51 $\pm$ 2 & 99 $\pm$ 5 & 36 $\pm$ 2 \\
	WR/(WR+O)          & 0.023 $\pm$ 0.007  &  0.023 $\pm$ 0.006 &  0.07 $\pm$ 0.03  & 0.004 $\pm$ 0.002 \\
	\noalign{\smallskip}    
  \hline
  \enddata
  \tablenotetext{a}{Assuming that 1 WNL star has a $L$(He II $\lambda$4886) = 1.7 $\times$ 10$^{36}$ erg s$^{-1}$ \citep{VC92}.}
  \tablenotetext{b}{Assuming that 1 O7V star has a $L$(\Hb) = 4.76 $\times$ 10$^{36}$ erg s$^{-1}$ \citep{VC92}.}
  \tablenotetext{c}{This parameter is defined by $\eta\equiv$O7V/O \citep{SV98} and depends of the age of the burst.}
\end{deluxetable}

Although previous authors have derived the age of the last star-formation burst of NGC 5253, we have taken advantage of the 
high quality of our VLT spectra to make a reassessment of it through the \Ha\ equivalent width, $W$(H$\alpha$). 
We have used STARBURST99 \citep{L99} spectral synthesis 
models to estimate the age of each region, comparing the predicted \Ha\ equi\-valent widths with the observed values. 
We consider a spectral synthesis model with a metallicity of $Z$=0.008, the appropriate for NGC 5253, 
assuming an instantaneous burst with a Salpeter IMF, a total stellar mass of 10$^6$ $M_\odot$ and a 100 $M_\odot$ upper 
stellar mass. The ages obtained are compiled in Table~\ref{WR}, and are in agreement with previous results.

In Figure~\ref{roii} we can observe the broad emission feature of 
\ion{He}{2} $\lambda$4686 originating from the stellar winds of WR stars, which is present in the four observed regions of NGC 5253. 
The \ion{He}{2} $\lambda$4686 emission line is the most prominent feature of the so-called blue WR bump. Unfortunately, 
the existence of the red WR bump, that basically consists on the \ion{C}{4} $\lambda$5808 emission line, 
could not be studied because this spectral region is in our observational gap between 5783 and 5830 \AA. In Figure~\ref{roii} 
it is also evident the presence of broad emission features of \ion{N}{3} $\lambda$4634 and \ion{N}{3} $\lambda$4640, 
which are characteristic of WNL stars \citep*{SSM96}.

\begin{figure}[t!]
\includegraphics[angle=270,width=\linewidth]{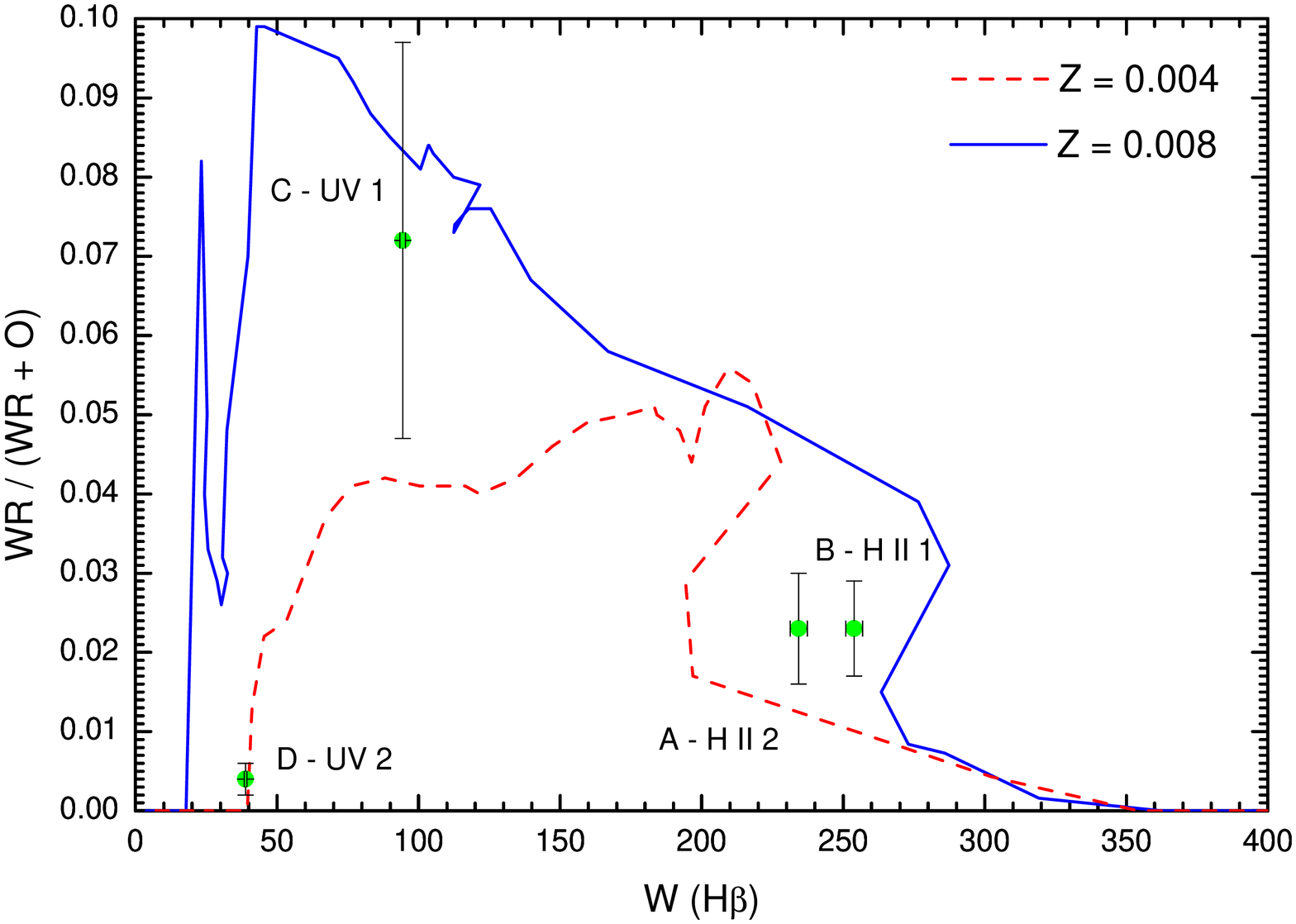}
\caption{\small{WR/(WR+O) versus W(H$\beta$) for \citet{SV98} models compared with our derived ratios.}}
\label{wr}
\end{figure}

We have used evolutionary synthesis models by \citet{SV98} for O and WR populations in young starburst to estimate 
the number of O and WR stars in the regions using the method explained in \citet*{LSER04a}. We have assumed that all the flux 
of the blue WR bump comes from the broad \ion{He}{2} $\lambda$4686 emission line (we show its extinction-corrected flux 
in  Table~\ref{WR}) and a distance of 3.3 Mpc for NGC 5253. We only need 1 WNL star 
in regions A and B to produce the observed WR spectral features. The number 
of O stars was derived using the procedure given 
by \citet*{GIT00} (their eq. 6). The derived number of O total stars and the WR/(WR+O) ratio are 
quoted in Table~\ref{WR}. 

We have found a very good correspondence between our results and the predictions from \citet{SV98} models. In Figure~\ref{wr} we plot the WR/(WR+O) 
ratio versus $W$(\Hb) for the models with $Z$=0.008 and 0.004. The metallicity of the analyzed regions inside NGC 5253 is $Z\sim$0.006 
for A and B and $\sim$0.008 for C and D. 

\section{The localized nitrogen enrichment}

As it was commented in \S 1, several authors have reported localized nitrogen enrichment in areas coincident with our 
regions A and B. \citet{WR89} and \citet{Kobulnicky97} proposed that this 
localized enhancement is due to pollution by the winds of massive stars.  However, for 
\citet{Kobulnicky97} a difficulty of this hypothesis was the apparent lack of accompanying He enrichment in the zones 
where the N/H ratio is high. This fact was in contradiction with the observational and theoretical understanding of 
massive star yields. In this sense, the slightly higher He/H ratio we find in zones A and B with respect to C can solve 
somehow this puzzle. However, it must be taken into account that the detection of a modest helium pollution is difficult due to 
several reasons: a) the initial 
abundance of this element is far much higher than nitrogen, so the relative enrichment should be 
strong to be unambiguously detected; b) the uncertainty introduced by the assumed ICF to correct for the presence of 
neutral helium.

Our spectroscopic data include many more \ion{He}{1} lines than previous works, our measurements have much higher signal-to-noise ratio,
and we are using a more detailed method for determining the He$^+$ abundance. For example, \citet{Kobulnicky97} use a 
single line: \ion{He}{1} $\lambda$5876, to determine the He$^+$/H$^+$ ratio. The intensity of that line is rather  
affected by collision and radiation transfer effects, and the calculations for correcting for these effects have improved since 
Kobulnicky et al. paper. In any case, the possible localized helium enrichment we detect is only marginal considering the quoted 
uncertainties. Unfortunately, it is clear that any result is necessarily unconclusive. 

We have made a rough estimation of the mass of newly-created helium and nitrogen (stellar yield) necessary to produce the observed 
overabundances in regions A and B. We have reproduced the calculations of \citet{Kobulnicky97} for obtaining the total 
ionizing mass of regions A and B, assuming the same angular size and filling factor, but considering a distance of 3.3 Mpc instead of 
the 4.1 Mpc assumed by those authors. 
In Table~\ref{yields} we show the values we obtain for the cases of considering or not temperature fluctuations in the abundances, although 
the difference is almost irrelevant. We have considered the mean He and N abundances of regions A and B, and the yields are 
computed relative to the abundance of those elements in region C. Therefore, we have assumed that the initial abundance of NGC 5253 is that 
measured in region C and that zones A and B have suffered a localized and very recent increment of the He/H and N/H ratios. 
In Table~\ref{yields} we have compared our empirical stellar yields with those 
obtained by \citet{MM02} in their stellar evolution models including rotation. From the table, it can be seen that the contribution of 
a few evolved (WR) massive stars is enough to produce the observed pollution, in agreement with the low numbers of WNL stars estimated 
in \S 11 for zones A and B from the flux of the blue WR bump. The comparison with the also empirical yields determined 
by \citet{E92} for ring nebulae associated to Galactic WR stars gives further consistency to the hypothesis of the 
pollution by winds of WR stars. In particular, the similarity of the ratio of the He and N stellar yields estimated for NGC 5253 and for the other 
objects shown in Table~\ref{yields} is the most remarkable fact of the table. That ratio is independent of the quite uncertain assumptions considered 
for deriving the empirical yields of the individual elements. 

As it was commented in \S 8 and can be noted in Table~\ref{totabun}, the O/H ratio in zone C 
is marginally higher that in zones A and B, (although the values can be considered similar taking into account the errors). It is interesting to note 
that nebular ejecta around WR stars \citep{E92} and LBV stars \citep{smith97,lamers01} show a substantial oxygen deficiency in their chemical 
content. This is also consistent with the scenario of pollution by evolved massive stars.

The presence of localized chemical enrichment in the youngest starburst of NGC 5253 indicates that the timescale of the process should be very short, 
as it was also suggested by \citet{Pustilnik04}. The similarity of the enrichment pattern observed in this galaxy and that 
of the Galactic WR ring nebulae (with estimated lifetimes of the order of 10$^4$--10$^5$ years, see Esteban et al. 1992) is another piece of 
evidence in the same direction. The most suitable scenario is that the pollution process is produced by the ejection of chemically 
enriched  
stellar external layers which are photoionized by the surrounding massive stars. This is, for example, the formation mechanism of the 
nebulae around luminous blue variables (LBVs), which also show a similar N enhancement \citep{smith97,lamers01}. In fact, \citet{lamers01} propose 
that 
the LBV nebulae are ejected during the blue supergiant phase of the progenitor star and that the chemical enrichment is due to rotation-induced 
mixing. 
Conversely, the observed pollution is rather unlikely to be produced by stellar wind material confined within a hot bubble. In this case, the 
timescale for mixing would be far much larger, of the order of $\sim$100 Myr \citep{TT96} because it should cool-down to temperatures of the 
order of the ionized gas.

\begin{deluxetable}{l c@{\hspace{10pt}} c@{\hspace{10pt}} c@{\hspace{10pt}}  
c@{\hspace{10pt}} c@{\hspace{10pt}} c@{\hspace{10pt}}}
\tabletypesize{\footnotesize}
\tablecaption{Comparison of empirical and theoretical stellar yields$^{\rm a}$\label{yields}}
\tablewidth{0pt}
\tablehead{
\colhead{} &
\multicolumn{2}{c}{} &
\multicolumn{3}{c}{Contribution of stellar winds$^{\rm b}$} &
\colhead{} 
\\
\colhead{} &
\multicolumn{2}{c}{NGC 5253 (A and B)} & 
\multicolumn{2}{c}{40 M$_\odot$} &
\colhead{60 M$_\odot$} &
\colhead{Galactic WR} 
\\
\colhead{Stellar yield}   & 
\colhead{($t^2$=0)}  &
\colhead{($t^2$$>$0)}  & 
\colhead{($v_{\rm rot}$=0)} &
\colhead{($v_{\rm rot}$=300)} &
\colhead{($v_{\rm rot}$=0)} &
\colhead{Ring Nebulae$^{\rm c}$}  } 
\startdata
$mp_{\rm N}$ & 0.03 & 0.04 & 0.002& 0.026& 0.015& 0.002$-$0.02 \\
$mp_{\rm He}$ & 5 & 7 & 0.2 & 4.1 & 2.1 & 0.2$-$3.6 \\
$mp_{\rm He}$/$mp_{\rm N}$ & 167 & 175 & 100 & 158 & 140 & 90$-$180 \\    
	\noalign{\smallskip}    
	\hline
  \enddata
\tablenotetext{a}{Values given in solar masses.}
\tablenotetext{b}{Assuming $Z$/$Z_{\odot}$=0.2 and different initial stellar rotation velocities (in km s$^{-1}$) \citep{MM02}.}
\tablenotetext{c}{Empirical yields derived for a sample of Galactic ring nebulae around WN stars \citep{E92}.}
\end{deluxetable}

\section{About the origin of the abundance \\ discrepancy in NGC 5253}

Very recently, \citet{Tsamis05} have proposed an explanation for the abundance discrepancy based on the implicit acceptation of the 
existence of temperature fluctuations but produced by a chemically inhomogeneous medium. These authors propose a chemically inhomogeneous model for 
30 Doradus of low temperature and high metallicity 
embedded in a lower density medium of high temperature and low metallicity. The dense high metallicity regions come from Type II SN 
material that has not been mixed with the bulk of ISM and is in pressure balance with the normal composition 
ambient gas. These inclusions would be responsible for most of the emission 
in RLs with virtually no emission in CELs due to their very low electron temperature. We will present some objections to their models below.

\begin{deluxetable}{l c@{\hspace{10pt}} c@{\hspace{10pt}} c@{\hspace{10pt}} c@{\hspace{10pt}} c@{\hspace{10pt}} c@{\hspace{10pt}}  
c@{\hspace{10pt}} c@{\hspace{10pt}} }
\tabletypesize{\footnotesize}
\tablecaption{Comparison of the abundance discrepancy factor in different HII regions\label{adf}}
\tablewidth{0pt}
\tablehead{
\colhead{} &
\colhead{} &
\colhead{Galaxy} &
\colhead{} &
\colhead{$R_G$$^{\rm a}$} &
\colhead{} &
\colhead{ADF(O$^{++}$)$^{\rm c}$} &
\colhead{ADF(C$^{++}$)$^{\rm d}$} &
\colhead{} \\
\colhead{Object} &
\colhead{Galaxy} &
\colhead{Type} &
\colhead{M$_{\rm V}$} &
\colhead{(kpc)} &
\colhead{12+log(O/H)$^{\rm b}$} &
\colhead{(dex)} & 
\colhead{(dex)} &
\colhead{Reference$^{\rm e}$}} 
\startdata
M 16         & Milky Way & Spiral    & $-$20.9 & 6.34    & 8.56 & 0.45        & \nodata        & 1 \\ 
M 8          &           &           &         & 6.41    & 8.51 & 0.37        & 0.35           & 2 \\
M 17         &           &           &         & 6.75    & 8.52 & 0.27        & \nodata        & 2 \\
             &           &           &         &         & 8.56 & 0.32        & \nodata        & 3 \\
M 20         &           &           &         & 7.19    & 8.53 & 0.33        & \nodata        & 1 \\
NGC 3576     &           &           &         & 7.46    & 8.56 & 0.24        & \nodata        & 4 \\
             &           &           &         &         & 8.52 & 0.32        & \nodata        & 3 \\
Orion nebula &           &           &         & 8.40    & 8.51 & 0.14        & 0.40           & 5 \\
             &           &           &         &         & 8.52 & 0.11        & 0.38           & 3 \\
NGC 3603     &           &           &         & 8.65    & 8.46 & 0.29        & \nodata        & 1 \\
S 311        &           &           &         & 10.43   & 8.39 & 0.27        & \nodata        & 6 \\
NGC 5461     & M 101     & Spiral    & $-$21.6 & \nodata & 8.56 & 0.29        & $-$0.03/0.34   & 7 \\
NGC 604      & M 33      & Spiral    & $-$18.9 & \nodata & 8.49 & 0.20        & \nodata        & 7 \\
30 Doradus   & LMC       & Irregular & $-$18.5 & \nodata & 8.33 & 0.21        & 0.25           & 8 \\
             &           &           &         &         & 8.34 & 0.30$-$0.43$^{\rm f}$ & 0.41           & 3 \\
N11B         &           &           &         & \nodata & 8.41 & 0.69$-$0.91$^{\rm f}$ & \nodata        & 3 \\
N66          & SMC       & Irregular & $-$16.2 & \nodata & 8.11 & 0.36        & \nodata        & 3 \\
Region V     & NGC 6822  & Irregular & $-$16.0 & \nodata & 8.08 & 0.29        & \nodata        & 9 \\
NGC 2363     & NGC 2366  & Irregular & $-$18.2 & \nodata & 7.87 & 0.34        & 0.31:          & 7 \\
Zones A and B& NGC 5253  & BCDG      & $-$17.2 & \nodata & 8.18 & 0.23        & 0.41$^{\rm g}$ & 10 \\  
	\noalign{\smallskip}    
	\hline
  \enddata
\tablenotetext{a}{Only for Milky Way objects.}
\tablenotetext{b}{Value determined from CELs.}
\tablenotetext{c}{Defined as log(O$^{++}$/H$^+$)(RLs) $-$ log(O$^{++}$/H$^+$)(CELs).}
\tablenotetext{d}{Defined as log(C$^{++}$/H$^+$)(RLs) $-$ log(C$^{++}$/H$^+$)(UV CELs).}
\tablenotetext{e}{1- Garc\'\i a-Rojas et al. (2006); 2- Garc\'\i a-Rojas et al. (in preparation); 3- Tsamis et al. (2003); 4- Garc\'\i a-Rojas et al. 
(2004); 5- Esteban et al. (2004); 6- Garc\'\i a-Rojas et al. (2005);  7- Esteban et al. (2002); 8- Peimbert (2003); 9- Peimbert et al. (2005); 10- 
This work.}
\tablenotetext{f}{The highest values of the ADFs include a correction for underlying scattered light.}
\tablenotetext{g}{Considering only zone A.}
\end{deluxetable}

First, the values of the abundance discrepancy factor (ADF) of O$^{++}$ (defined as the difference between the O$^{++}$ abundance determined from RLs 
and CELs) found for very different HII regions in different hosts galaxies are rather similar, as it can be seen in Table~\ref{adf}. 
In this table we show the ADF values determined for a sample of Galactic and extragalactic \ion{H}{2} regions where the ADF of O$^{++}$ has 
been determined, their O/H ratio, the morphological type and absolute magnitude 
of the host galaxy and the Galactocentric distance in the case of the Galactic nebulae. It can be seen that the ADF is similar for
most of the objects shown in the table, independently of the mass, type and metallicity of their host galaxy. The 
galaxies included in Table~\ref{adf} have very different metallicities and gravitational potential wells and they should also have very different 
star formation histories and 
SN rates. In particular, in the case of NGC 5253, a high likelihood of a blow-out has been recently postulated by \citet{Summers04}. 
The energy injection 
rate into this galaxy due to the mechanical action of massive stars would seem sufficient to allow the expanding hot gas to escape its gravitational 
potential well. Moreover, the relatively small \ion{H}{1} halo ($\sim$1.8 kpc along the minor axis) of this dwarf galaxy would also offer little 
resistance to this blow-out. The likelihood of past blow-outs suggests that the proposed delayed enrichment mechanism based on the rain 
of cooled-down droplets of SN ejecta would be less efficient in this galaxy, but this is not reflected in a lower value of the ADFs 
in NGC 5253 with respect to the other \ion{H}{2} regions inside more massive galaxies. On the other hand, 
the Galactic \ion{H}{2} regions studied in our sample are located at different galactocentric distances (from 6 to 11 kpc) in the 
Galactic disk. If the hypothesis of the enrichment by H-poor inclusions of SNe ejecta is correct, this would imply that 
this process is rather independent of the conditions and properties of the Galactic disk in such interval or radial distances. 

Another important piece of evidence in this discussion is the fact that the ADF
of C$^{++}$ in NGC 5253 (and other objects where this quantity has been 
determined, see Table~\ref{adf}) is rather similar to that of oxygen. We do not
see any metallicity dependence of the relative importance of the ADF of 
C$^{++}$ with respect to 
that of O$^{++}$ in the objects where both quantities have been estimated (see
Table~\ref{adf}). 
According to the model by \citet{Tsamis05}, the overabundances in the metal rich
regions amount to a factor of 8 in O and a factor of 14 in C. It is well 
known that massive star evolution mo\-dels predict 
(e.g. Maeder 1992; Portinari, Chiosi, \& Bressan 1998) that most of the net C is
produced before the onset of the SN stage via stellar winds, and 
since the wind strength increases with metallicity, the C yield increases with
metallicity. 
\citet*{Carigi06} have produced a chemical evolution model for NGC 6822, a dwarf
irregular galaxy of the Local Group, that predicts that only 36\% of the 
C is due to massive stars that produce SN of type II, while 63\% of the C is due
to low and intermediate mass stars that do not produce Type II SN and, 
therefore, do not contribute to the C content of the droplets. 
\citet{Matteucci06} also finds that for the solar neighborhood most of the C is
due to low and intermediate mass stars, while \citet{Carigi05} find 
that about half of the C comes from intermediate mass stars and half from
massive stars. Considering that the 30 Doradus chemical composition is similar
to that of NGC 6822 V (see Table~\ref{adf}) one can reasonably infer that the
expected overabundance of C should be lower than that of O, contrary to the 
assumptions of \citet{Tsamis05}. Similarly, in the chemical evolution models
most of the N is produced by low and 
intermediate mass stars and not by massive stars, consequently a considerable
smaller overabundance of N than O is expected, contrary to the N and O values
adopted in the model by \citet{Tsamis05}.

\section{Conclusions}

We present deep echelle spectrophotometry in the 3100$-$10400 \AA\ range of four selected zones of the nucleus of the blue compact dwarf galaxy 
NGC~5253. We have identified and measured a large number of emission lines in the spectra. This represents the largest collection of optical emission 
lines available for a dwarf starburst galaxy. Electron densities and temperatures have been consistently determined from a large number of emission  
line ratios of different ions. Chemical abundances of He, N, O, Ne, S, Cl, Ar, and Fe have been derived following standard methods. Recombination 
lines of \ion{C}{2} and \ion{O}{2} have been measured in the spectra of three of the individual zones analysed. The first time these kinds of 
lines have been unambiguously detected in a dwarf \ion{H}{2} galaxy. The ionic abundances of C$^{++}$ and O$^{++}$ derived from these lines are from 
0.20 to 0.40 dex larger than those determined from the intensity of collisionally excited lines. This behavior has been found in all the Galactic and 
extragalactic \ion{H}{2} regions where both kinds of lines have been reported for the same ion. The quoted abundance differences can be accounted for 
a temperature fluctuations parameter ($t^2$) between 0.050 and 0.072. 

The emission line profiles are complex and seem to be the combination of $-$at least$-$ two kinematical components with slightly different radial 
velocity and, in some zones, different line width. Apparently, there are no clear differences in the physical and chemical properties of the 
kinematical components, except perhaps a somewhat different reddening coefficient and N/O ratio in some particular zones. Position-velocity diagrams 
of different emission lines show a sinusoidal shape. However, the position-velocity diagram of stellar absorption spectral features shows a pattern 
entirely consistent with rotation. This fact suggests that the radial velocity behavor of the ionized gas should be due to outflows from the central 
starbursts and not a product of merging.  We have derived the ages and the massive star population of the starburst. In particular, the detection of 
the so-called blue WR bump indicates the presence of a rather small number of WR stars in the starbursts. 

Our observations confirm previous results that indicate the presence of localized N enrichment in two of the central starbursts of NGC~5253. 
Moreover, 
we also detect a possible slight He pollution in the same zones. The enrichment pattern is entirely consistent with that expected by the pollution of 
the winds of masive stars in the WR phase. Moreover, the estimated mass of newly created N and He is also consistent with the number of WR stars 
determined in the starbursts. 

Finally, we conclude that a recent hypothesis that tried to solve the abundance discrepancy problem in \ion{H}{2} regions 
in terms of a delayed enrichment by SNe ejecta inclusions \citep{Tsamis05} seems not to explain the available observational data in a 
satis\-factory manner.

\acknowledgments

This work is based on observations collected at the European Southern Observatory, Chile, proposal number ESO 
70.C-0008(A). We are grateful to Mar\'{\i}a Teresa Ruiz for her collaboration in the initial stages of this project. We thank Conrado Carretero for 
his comments about CaT absorptions and B\"arbel Koribalski for share with us her preliminary results about new HI ATCA observations of NGC 5253. This 
work has been partially funded by the Spanish Ministerio de Ciencia y Tecnolog\'{\i}a (MCyT) under project 
AYA2004-07466. MP received partial support from DGAPA UNAM (grant IN114601). MR acknowledges support from Mexican CONACYT project J37680-E. This 
research has made use of the NASA/IPAC Extragalactic Database (NED) which is operated by the Jet Propulsion Laboratory, California Institute of 
Technology, under contract with the National Aeronautics and Space Administration.

\end{document}